\documentclass{IEEEtran}
\usepackage{cite}
\usepackage{amsmath,amssymb,amsfonts}
\usepackage{algorithmic}
\usepackage{graphicx}
\usepackage{multicol}
\usepackage{textcomp}
\usepackage[table]{xcolor}
\usepackage{setspace}
\usepackage{enumitem}
\usepackage{slashbox}
\def\BibTeX{{\rm B\kern-.05em{\sc i\kern-.025em b}\kern-.08em
		T\kern-.1667em\lower.7ex\hbox{E}\kern-.125emX}}
\usepackage{subfigure}

\newtheorem{my_lemma}{Lemma}

\newtheorem{my_proposition}{Proposition}

\newtheorem{my_definition}{Definition}
\title{Meijer-G Function with Continued Product and Integer Exponent: Performance of Multi-Aperture UOWC System over EGG Turbulence} 
\author{Arvind Kumar,~\IEEEmembership{Graduate Student Member,~IEEE}, Nikumani Choudhury, Jayendra N. Bandyopadhyay,   S.~M.~Zafaruddin,~\IEEEmembership{Senior Member,~IEEE} 
	\thanks{A preliminary version of this work containing an analysis of the average BER for i.i.d. channel has been presented at the 2024 IEEE International Black Sea Conference on Communications and Networking (BlackSeaCom), Tbilisi, Georgia, 24-27 June 2024  \cite{blackseacom_2024}.}
	\thanks{ This work was supported in part by the Anusandhan National Research Foundation (ANRF), Department of Science and Technology (DST), Government of India under CRG/2023/00840 and  MTR/2021/000890,  in part by BITS Pilani CDRF C1/23/120, and in part by the  TTDF (DoT, Govt. of India) under the Accelerated Research on 6G Ecosystem TTDF/6G/123.}	
	\thanks{Arvind Kumar (p20230404@pilani.bits-pilani.ac.in) and S.~M.~Zafaruddin (syed.zafaruddin@pilani.bits-pilani.ac.in)  are with the Department of Electrical and Electronics Engineering, Birla Institute of Technology and Science, Pilani, Pilani Campus,  Pilani-333031, Rajasthan, India.	}
	
	\thanks{Nikumani Choudhury (nikumani@hyderabad.bits-pilani.ac.in) is with the Department of Computer Science \& Information Systems, Birla Institute of Technology and Science, Pilani, Hyderabad Campus,  Hyderabad, India.	}
	
	\thanks{Jayendra N. Bandyopadhyay (jayendra@pilani.bits-pilani.ac.in) is with the Department of Physics, Birla Institute of Technology and Science, Pilani, Pilani Campus,  Pilani-333031, Rajasthan, India.		}

}

\begin{document}
	\maketitle
	\begin{abstract}
		Signal transmission over underwater optical wireless communication (UOWC) experiences the combined effect of oceanic turbulence and pointing errors statistically modeled using the sum of two Meijer-G functions. There is a research gap in the exact statistical analysis of multi-aperture UOWC systems that use selection combining diversity techniques to enhance performance compared to single-aperture systems. In this paper, we develop a general framework for the continued product and positive integer exponent for the sum of Meijer-G functions to analyze the exact statistical performance of the UOWC system in terms of multivariate Fox-H function for both independent and non-identically distributed (i.ni.d.) and independent and identically distributed (i.i.d.) channels. We also approximate the performance of a multi-aperture UOWC system with i.i.d.  channels using the single-variate Fox-H function. Using the generalized approach,   we present analytical expressions for average bit-error rate (BER)  and ergodic capacity for the considered system operating over exponential–generalized gamma (EGG) oceanic turbulence combined with zero-boresight pointing errors. We also develop asymptotic expressions for the average BER at a high signal-to-noise (SNR) to capture insights into the system's performance.  Our simulation findings confirm the accuracy of our derived expressions and illustrate the impact of turbulence parameters for i.ni.d. and i.i.d. models for the average BER and ergodic capacity, which may provide a better estimate for the efficient deployment of UOWC.
	\end{abstract}
	

	%%Utilizing a multi-aperture receiver with diversity combining techniques holds promise in addressing signal fading for UOWC transmission.
	\section{Introduction}
	Next generation 6G wireless system promises to extend ubiquitous connectivity for emerging applications, including non-terrestrial underwater communications \cite{dang2020what}. Underwater activities have gained significant attention due to their applications in environmental monitoring, oil and gas management, coastal security, and military underwater vehicles \cite{Gussen2016, kaushal2016underwater, zeng2016survey}. 
	Due to the limitations of conventional RF and acoustic carriers in underwater communication networks, including low data rates and significant delays, the research community has increasingly focused on alternatives like underwater optical wireless communication (UOWC), which offers high data speeds and compatibility with emerging 6G technologies. 
	UOWC displays potential for submarines and oceanic research applications due to its ability to attain increased data transmission rates, which can be achieved by using extensive optical bandwidth. Despite the numerous advantages of UOWC, underwater communication links face challenges such as signal attenuation, oceanic turbulence, and pointing errors. Signal attenuation arises from molecular absorption and photon scattering as light propagates through water, typically characterized by the extinction coefficient \cite{Lucas2023On}. Oceanic turbulence is caused by random fluctuations in the refractive index of the UOWC channel, resulting from variations in water temperature, salinity, and the presence of air bubbles. Further, pointing errors due to misalignment between the transmitter and receiver apertures can impact UOWC transmissions.
	
	Statistical channel impairments in UOWC systems encompass pointing errors and oceanic turbulence. The zero-boresight model for pointing errors \cite{Farid2007} has been adopted for UOWC to facilitate tractable performance analysis. Several statistical models for optical turbulence include log-normal, Gamma, $K$, Weibull, exponentiated Weibull distributions \cite{Vahid2016, Mingjian2016, Vahid2015,  Ahmadirad2018,Zhou2024}. The exponential-generalized Gamma (EGG) distribution showed strong agreement with data gathered across a spectrum of channel conditions, spanning from mild to strong turbulence \cite{Zedini2019}. The statistical distribution and density function for the combined channel with pointing errors and the EGG turbulence can be presented using the sum of two Meijer-G functions consisting of Melin-Barnes integral over multiple Gamma functions. The EGG model has attracted considerable focus in assessing the performance of UOWC systems concerning  UOWC link \cite{Zedini2019,Zedini2020_TCOM, Le2022,  Xiaobin2020}, integrated terrestrial-UOWC \cite{Naik2024multiuser,   yang2021performance, wang2024performance}, vertical cascaded UOWC \cite{Rahman2022_unified, lou2022performance}, optical reconfigurable intelligent surface (ORIS) \cite{Romdhane2023, Rakib2024ris}, and multi-user systems \cite{zhang2023performance, Rahman2024_capacity}.

	Multi-aperture UOWC can be a potential technique to minimize the channel impairment by harnessing the randomness in the signal reception \cite{zeng2016survey, boucouvalas2016underwater, liu2015simo, Jamali2017, Ansari2019,Bansal2023, Romdhane2023}. The authors in \cite {boucouvalas2016underwater, liu2015simo} analyzed spatial diversity schemes for weak oceanic turbulence modeled by the log-normal distribution. In \cite{Jamali2017} spatial diversity for multi-input multi-out (MIMO) UOWC was presented for log-normal turbulence. Authors in \cite{Ansari2019} investigated the outage probability of a mixed terrestrial OWC connection with multi-sensor UOWC under weak oceanic and atmospheric turbulence. Recently, the authors in \cite{Bansal2023, Romdhane2023} analyzed the performance of selection combining (SC) and maximal ratio combining (MRC) receivers for multi-aperture UOWC system under the combined effect of EGG oceanic turbulence and pointing errors. However, the analysis in \cite{Bansal2023} was limited to the outage probability, and for the i.i.d. model, the outage performance of the MRC  was presented using an upper bound. The authors in \cite{Romdhane2023} extended the outage analysis for the MRC for the i.ni.d. model using multivariate Fox-H. As a proof of concept, the authors in \cite{Weijie2023Experimental}  experimentally evaluated the performance of a multi-aperture receiver system in mitigating temperature-induced turbulence in a UOWC system with a correlated single-input multiple-output (SIMO) channel. Through controlled experiments, the multi-aperture setup employed the MRC to optimize signal reception by coherently integrating signals from multiple apertures,  leading to enhanced performance and reliability of the UOWC system under turbulent conditions. It should be emphasized that the SC diversity scheme is simple to implement without requiring channel state information (CSI) and performs close to the MRC for a reasonable number of apertures under various signal-to-noise ratios (SNRs), as demonstrated in \cite{Bansal2023}.

	\subsection{ Novelty  and Contributions}
	Based on the above and related research (see the following subsection), there is a gap in the statistical performance regarding the average bit-error rate (BER) and ergodic capacity for SC-based multi-aperture UOWC systems. Developing exact analytical expressions for the average BER and ergodic capacity poses challenges for both  i.ni.d. and i.i.d. multi-aperture UOWC systems equipped with the SC receiver. 
	The statistical analysis of the i.ni.d. UOWC system involves a continued product of the sum of two Meijer-G functions. In contrast, the i.i.d. system consists of Meijer-G functions with a positive integer exponent, necessitating novel approaches for performance evaluation. Although the performance of wireless systems (such as free-space optics (FSO),  terahertz (THz), and radio frequency (RF)) involving a single Meijer-G function has been extensively analyzed in the literature, analysis involving a sum of two Meijer-G functions for multi-aperture/multi-antenna has not been dealt with. Further, statistical analysis for the integer exponent, even for a single Meijer-G function for fading channels, has yet to be studied. It should be noted that the existing mathematical framework for the average BER and ergodic capacity can deal with the product of multiple Meijer-G functions but not the  Meijer-G function with an arbitrary integer exponent.

	This paper provides an accurate performance evaluation of an SC-based multi-aperture UOWC system using average BER and ergodic capacity to optimize deployment scenarios. The major contributions of the paper are as follows:
	
	\begin{itemize}
		\item We develop a novel approach to deal with the continued product and integer exponent for the sum of Meijer-G functions to analyze the exact statistical performance for both i.ni.d. and i.i.d. SC-based multi-aperture UOWC system in terms of multivariate Fox-H function.
		
		\item We approximate the performance of i.i.d. multi-aperture UOWC system using a single-variate Fox-H by exploiting the contours of multiple Mellin-Barnes integrals over the same integrand. This approach can be extended to analyze the performance of other wireless systems, such as FSO,  THz, and RF, involving a Meijer-G function with an integer exponent. 
		
		\item We derive analytical expressions for average BER  and ergodic capacity for  SC-based multi-aperture UOWC system under the combined effect of the EGG  oceanic turbulence and zero-boresight pointing errors in terms of multivariate and single variate Fox-H function considering various propagation scenarios.
		
		\item We validate the derived analytical expressions through Monte Carlo simulations considering parametric values for various oceanic turbulence conditions and illustrate that the exact expressions for the average BER and ergodic capacity may provide a better estimate for the efficient deployment of UOWC.
	\end{itemize}
	
	\subsection{Related Work}

	In recent years, extensive research has focused on using optical communications for underwater applications. Given the importance of statistical performance analysis, UOWC systems have been studied using several optical turbulence models, including the log-normal, Gamma, Gamma-Gamma, 
	$K$, Weibull, exponentiated Weibull, and Weibull–generalized gamma (WGG) distributions \cite{Vahid2015, Mingjian2016, Vahid2017, Ahmadirad2018, Zhou2024}. A study on relay-assisted UOWC with optical code division multiple access (OCDMA) over log-normal turbulent channels is presented in \cite{Vahid2016}. In \cite{Mingjian2016}, the authors explored analytical performance for the channel capacity of UOWC transmission under weak oceanic turbulence. The outage probability, average BER, and ergodic capacity for UOWC using the EGG channel were presented in \cite{Zedini2019}. The authors in \cite{Zhou2024} recently provided a statistical analysis for a RIS-assisted system considering WGG turbulence. Recent studies have also extended single-link models to cascaded models, considering the vertical inhomogeneity of the UOWC channel due to depth-related gradient effects.

	The use of emerging technologies such as RIS and multiple access techniques such as non-orthogonal multiple access (NOMA) and rate-splitting multiple access (RSMA) has been studied for UOWC \cite{Rakib2024ris,Salam2023optical,Salam2023dynamic,Rahman2024_capacity,zhang2023performance, li2024performance,Feng2024,sharma2023uplink}. A study by \cite{Rakib2024ris} carried out a performance evaluation of a new RIS-assisted, variable-gain dual-hop AF mixed THz-UOWC system considering the impact of pointing error. \cite{Salam2023dynamic} proposed mirror element assignment techniques for non-line-of-sight (NLOS) OIRS-assisted UOWC links for supporting multiple users.  \cite{Salam2023optical}  analyzed performance of mirror based OIRS and planar-mirror-surface (PMS) assisted UOWC system.  In \cite{Rahman2024_capacity}, the performance of RSMA-enabled multi-user UOWC systems was presented. A non-orthogonal multiple access (NOMA)-assisted dual-hop UOWC-RF network was studied in \cite{zhang2023performance, li2024performance}. In \cite{Feng2024}, authors analyzed the outage performance of a multi-antenna NOMA RF-UOWC system considering imperfect CSI, successive-interference-cancellation (SIC), and other system parameters such as the number of antennas, temperature and salinity of the water. The authors in \cite{sharma2023uplink} studied the outage performance of an uplink decode-and-forward (DF) mixed UOWC-RF system that employs NOMA for underwater transmission.

	Despite the increase in research on UOWC, exact analytical results for average BER and ergodic capacity in multi-aperture UOWC systems over EGG channels with positioning errors have not been studied. Analytical results for average BER and ergodic capacity are crucial for fine-tuning system parameters in practical scenarios involving underwater turbulence and pointing errors.
	
	\subsection{Organization of Paper}
	The paper is organized as follows: Section II describes the system and channel models. Section III addresses a generalized approach for analyzing the Meijer-G Function with continued products and integer exponents. Section IV presents the statistical performance regarding average BER and ergodic capacity for SC-based multi-aperture UOWC systems. Section V provides simulation and numerical analysis using MATLAB. Finally, Section VI concludes the paper.
	
	\section{System Model}
	\begin{figure}[tp]
		\centering
		\includegraphics[scale=0.8]{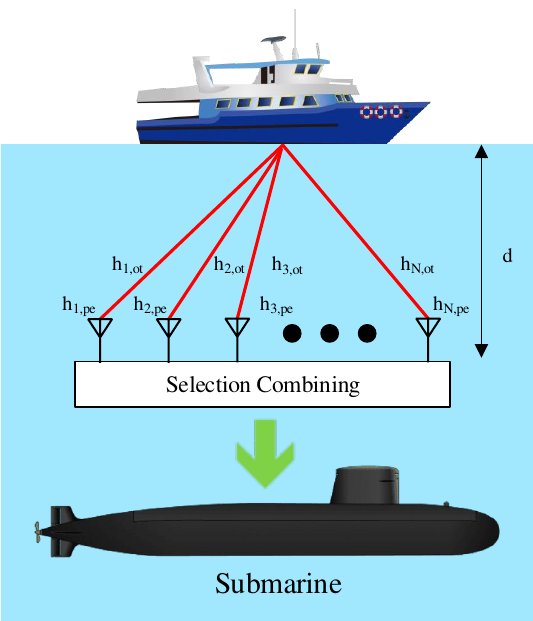}
		\caption{A schematic diagram of the multi-aperture UOWC system.}
		\label{model_diag}
	\end{figure}
	We consider a UOWC system where a link is established from a buoy to a user located a distance $d$  in the sea. Since a single aperture may not provide acceptable performance under the combined effect of signal attenuation, oceanic turbulence,  and pointing errors, we employ $N (\geq 1)$-aperture at the receiver to leverage the SC-based diversity from multiple signals, as shown in Fig.~\ref{model_diag}.

	Using on-off keying (OOK) and non-coherent intensity modulation/direct detection (IM/DD) scheme,  the instantaneous SNR at the $i$-th aperture  can be expressed as \cite{Farid2007} 
	\begin{equation}
		\gamma_i =  \bar{\gamma} h_{i, ot}^2h_{i, pe}^2
		\label{eq:inst.snr}
	\end{equation}
	where $\bar{\gamma}$ is the SNR without fading (taken same at all apertures since the apertures are close to each other),  and $ h_{i,ot}$ and $h_{i,pe}$ are channel coefficients for oceanic turbulence and pointing errors at the $i$-th aperture, respectively. 
	
	Assuming  EGG distribution for the oceanic turbulence \cite{Zedini2019} and zero-boresight pointing errors \cite{ijeh2022outage}, the probability density function (PDF) and cumulative distribution function (CDF) of the SNR at the $i$-th aperture in \eqref{eq:inst.snr}  can be expressed as \cite{yang2021performance}:
	\begin{eqnarray}
		&f_{\gamma_i}(\gamma) =  \frac{\omega_i \rho_i^2 }{2 \gamma} G_{1,2}^{2,0}\left(\begin{array}{c}\rho_i^2+1\\1,\rho_i^2\end{array}\left|\frac{1}{\lambda_i A_i} \left(\sqrt{\frac{\gamma}{\bar{\gamma}}}\right)\right.\right) \nonumber \\ & + \frac{(1-\omega_i)\rho_i^2}{2 \Gamma(a_i) \gamma} G_{1,2}^{2,0}\left(\begin{array}{c}\frac{\rho_i^2}{c_i}+1\\a_i,\frac{\rho_i^2}{c_i}\end{array}\left|\frac{1}{b_i^{c_i} A_i^{c_i}}\left(\sqrt{\frac{\gamma}{\bar{\gamma}}}\right)^{c_i}\right.\right),
		\label{pdf_uw_pe}
	\end{eqnarray}
	
	\begin{eqnarray}
		&F_{\gamma_i}(\gamma) = \omega_i\rho_i^2 G_{2,3}^{2,1}\left(\begin{array}{c}1,\rho_i^2+1\\1,\rho_i^2,0\end{array}\left|\frac{1}{\lambda_i A_i}\left(\sqrt{\frac{\gamma}{\bar{\gamma}}}\right)\right.\right) \nonumber\\&+\frac{(1-\omega_i)\rho_i^2}{c_i \Gamma(a_i)}G_{2,3}^{2,1}\left(\begin{array}{c}1,\frac{\rho_i^2}{c_i}+1\\a_i,\frac{\rho_i^2}{c_i},0\end{array}\left|\frac{1}{b_i^{c_i} A_i^{c_i}}\left(\sqrt{\frac{\gamma}{\bar{\gamma}}}\right)^{c_i}\right.\right),
		\label{cdf_uw_pe}
	\end{eqnarray}
	where  $\lambda_i$, $a_i$, $b_i$, and $c_i$ are the exponential distribution parameters, and $\omega_o$ is a mixture coefficient that determines the relative contribution of each component distribution to the overall distribution. Here, $A_i$ and $\rho_i$ are pointing error parameters at the $i$-th aperture \cite{ijeh2022outage}.
	
	The selection diversity scheme opportunistically selects a single signal with the best SNR $\gamma$ from   $N$ laser detectors in the receiver side of a multi-aperture UOWC system as $\gamma=\max \{\gamma_i\}_{i=1}^N$. We consider both i.ni.d. and i.i.d. channel models for statistical performance analysis. Applying the theory of random variables, the CDF  $F_{\gamma_{\rm inid}}({\gamma})$	for the i.ni.d. model  is given by 
	\begin{eqnarray}\label{cdf_sc_inid}
		F_{\gamma_{\rm inid}}({\gamma}) = \prod_{i=1}^NF_{\gamma_i}(\gamma)
	\end{eqnarray}
	Taking the derivative of  \eqref{cdf_sc_inid}, the PDF for the i.ni.d. model is represented as
	\begin{eqnarray}\label{pdf_sc_inid}
		f_{\gamma_{\rm inid}}({\gamma}) = \sum_{j = 1}^N \left( \prod_{i=1, i \neq j}^N F_{\gamma_i}(\gamma) \right)             f_{\gamma_j}({\gamma})
	\end{eqnarray}
	
	For the case of i.i.d. $F_{\gamma_1}(\gamma)=F_{\gamma_2}(\gamma)=F_{\gamma_3}(\gamma),\cdots, F_{\gamma_N}(\gamma)$, the CDF in \eqref{cdf_sc_inid} can be simplified as:
	\begin{eqnarray}\label{cdf_sc_iid}
		F_{\gamma_{\rm iid}}({\gamma}) = (F_\gamma({\gamma}))^N \label{cdf}.
	\end{eqnarray}
	Thus, the PDF for the i.i.d. model becomes
	\begin{eqnarray}\label{pdf_sc_iid}
		f_{\gamma_{\rm iid}}({\gamma}) = N(F_\gamma({\gamma}))^{N-1}f_\gamma({\gamma})
	\end{eqnarray}
	It is evident from \eqref{cdf_sc_inid} and \eqref{pdf_sc_inid} that performance analysis involving the CDF (used for the average BER) and PDF (used for ergodic capacity) necessitates handling a continued product of Meijer-G functions. Moreover, \eqref{cdf_sc_iid} and \eqref{pdf_sc_iid} illustrate that the performance analysis of the SC-based receiver involves the sum of Meijer-G functions with integer exponents. Consequently, analysis that requires a continued product of Meijer-G functions and Meijer-G functions with integer exponents becomes challenging and demands novel approaches.

	\section{Meijer-G Function with Continued Products and Integer Exponent}
	
	In this section, we develop a general framework for statistical analysis containing the Meijer-G function with continued product and integer exponent. To proceed, we require a simplified representation of Meijer-G and Fox-H using Mellin-Barnes integrals: 
	\begin{my_definition}
		The  Meijer-G function is defined as \cite{mathai2009h} 
		\begin{flalign}\label{eq:mg_def}
			G_{p,q}^{m,n}  \left(\begin{array}{c} \{a_{w}\}_{w=1}^{p}\\ \{b_{w}\}_{w=1}^{q} \end{array} \middle\vert \psi x \right)=\frac{1}{2\pi j}\int_L F(s) (\psi x)^s ds,
		\end{flalign}
		where $L$ is the contour integral, and   $F(s)$ is represented using Gamma function involving $s$ with coefficients $\{a_{w}\}_{w=1}^{p}$ and $\{b_{w}\}_{w=1}^{p}$.
	\end{my_definition}
	\begin{my_definition}
		The  Fox-H function is defined as \cite{mathai2009h} 
		\begin{flalign}\label{eq:h_def}
			H_{p,q}^{m,n}  \left( \begin{array}{c} \{(a_{w},A_{w})\}_{w=1}^{p}\\ \{(b_{w},B_{w})\}_{w=1}^{q} \end{array}  \middle\vert \psi x \right)=\frac{1}{2\pi j}\int_L F(s) (\psi x)^s ds,
		\end{flalign}
		where $L$ is the contour integral, and   $F(s)$ is represented using Gamma function involving $s$ with coefficients $\{a_{w}, A_w\}_{w=1}^{p}$ and $\{b_{w}, B_w\}_{w=1}^{p}$.
	\end{my_definition}

	The statistical performance of the SC-based UOWC system for i.ni.d. channels   may involve an integral of the following kind:
	\begin{eqnarray}\label{eq:IN1}
		I = \int_{0}^{\infty} \phi(x) \prod_{i=1}^N\bigg [ G_{p_i,q_i}^{m_i,n_i}  \left(\begin{array}{c} \{a_{i,w}\}_{w=1}^{p_i}\\ \{b_{i,w}\}_{w=1}^{q} \end{array} \middle\vert \psi_ix \right)\nonumber \\+G_{p'_i,q'_i}^{m'_i,n'_i}  \left(\begin{array}{c} \{a'_{i,w}\}_{w=1}^{p'_i}\\ \{b'_{i,w}\}_{w=1}^{q'} \end{array} \middle\vert \psi'_ix \right)\bigg] dx
	\end{eqnarray}
	where  $\phi(x)$ is a function of $x$. Extensive research has addressed the integration of the form in \eqref{eq:IN1}, but it is limited to a single Meijer-G function. In the following Proposition, we represent \eqref{eq:IN1} in a suitable form for analysis:
	\begin{my_proposition}\label{contproduct_sum}
		The integral in \eqref{eq:IN1} can be represented as
		\begin{flalign}\label{eq:IN12}
			&I = \sum_{S\in\{1,2, \cdots,N\}} \int_{0}^{\infty} \phi(x) \bigg[ \prod_{i\in S} G_{p_i,q_i}^{m_i,n_i}  \left(\begin{array}{c} \{a_{i,w}\}_{w=1}^{p_i}\\ \{b_{i,w}\}_{w=1}^{q} \end{array} \middle\vert \psi_ix \right)  \nonumber \\ & \times \prod_{i\in S^c} G_{p'_i,q'_i}^{m'_i,n'_i}  \left(\begin{array}{c} \{a'_{i,w}\}_{w=1}^{p'_i}\\ \{b'_{i,w}\}_{w=1}^{q'} \end{array} \middle\vert \psi_ix \right) \bigg]dx
		\end{flalign}
		where  $S \subseteq \{1,2,\cdots, N\}$ and $S^c$ is complimentary set of $S$ such that 	$S+S^c = \{1,2,3,\cdots, N\}$.
	\end{my_proposition}
	\begin{IEEEproof}
		From \eqref{eq:IN1}, we denote the first Meijer-G function as $g_i(x)$ and the second Meijer-G function as $g_i'(x)$:
		\begin{flalign}\label{eq:meijer_pq}
			&	g_i(x)=G_{p_i,q_i}^{m_i,n_i}  \left(\begin{array}{c} \{a_{i,w}\}_{w=1}^{p_i}\\ \{b_{i,w}\}_{w=1}^{q} \end{array} \middle\vert \psi_ix \right)\nonumber\\&
			g_i'(x)=G_{p'_i,q'_i}^{m'_i,n'_i}  \left(\begin{array}{c} \{a'_{i,w}\}_{w=1}^{p'_i}\\ \{b'_{i,w}\}_{w=1}^{q'} \end{array} \middle\vert \psi'_ix \right)
		\end{flalign}
		
		Validation of \eqref {eq:IN12} with $N=1$ is a trivial case. If $N=2$, we get
		\begin{flalign}\label{eq:cp1}
			&\prod_{i=1}^{2} (g_i(x) + g_i'(x)) = \prod_{i=1}^{2} g_i(x) + \prod_{i=1}^{2} g_i'(x)+ g_1(x)g_2'(x)\nonumber \\& +g_2(x)g_1'(x) \nonumber \\&
			=\sum_{S \subseteq \{1,2\}} \prod_{i\in S}g_i(x) \prod_{i\in S^{c}} g_i'(x)
		\end{flalign}
		
		Applying the iterative procedure,  we can express
		\begin{flalign}\label{eq:cp7}
			&	\prod_{i=1}^{N} (g_i(x) + g_i'(x)) = \sum_{S \subseteq \{1,2,.., N\}} \prod_{i\in S}g_i(x) \prod_{i\in S^{c}} g_i'(x)
		\end{flalign}
		We can use the induction method to verify  \eqref{eq:cp7}.	If \eqref{eq:cp1} holds true for $N = k$, we multiply \eqref{eq:cp7} by $g_{k+1}(x)+g'_{k+1}(x)$ to get
		\begin{flalign}\label{eq:cp5}
			&[g_{k+1}(x)+g'_{k+1}(x)]	\prod_{i=1}^{k} (g_i(x) + g_i'(x)) \nonumber \\ &  = \sum_{S \subseteq \{1,2,.., k\}} g_{k+1}(x)\prod_{i\in S}g_i(x)  +  \sum_{S \subseteq \{1,2,.., k\}} g'_{k+1}(x) \prod_{i\in S^{c}} g_i'(x)
		\end{flalign}
		Equation \eqref{eq:cp5} can be simplified to get 
		\begin{flalign}
			&\prod_{i=1}^{k+1} (g_i(x) + g_i'(x)) = \sum_{S \subseteq \{1,2,.., k+1\}} \prod_{i\in S}g_i(x) \prod_{i\in S^{c}} g_i'(x)
		\end{flalign}
		which is exactly same as \eqref{eq:IN12} with $N=k+1$, and thus proves the Proposition.
	\end{IEEEproof}
	The presentation in \eqref{eq:IN12} enables the straightforward application of the multivariate Fox-H function  \cite{mathai2009h}  to get a closed-form expression. In the next section, we use the representation in \eqref{eq:IN12} to derive analytical expressions of the average BER and ergodic capacity.

	It should be noted that performance analysis for i.i.d. channels is conducted as a special case of the i.ni.d. channel. Interestingly, direct performance analysis for  i.i.d. channels may involve the integral of  the following kind: 
	\begin{flalign}\label{eq:I16}
		I=\int_{0}^{\infty} \phi(x) \left(G_{p,q}^{m,n}  \left(\begin{array}{c} \{a_{w}\}_{w=1}^{p}\\ \{b_{w}\}_{w=1}^{q} \end{array} \middle\vert \psi x \right)\right)^M dx 
	\end{flalign}
	where $M$ is a positive integer. Although the formulation in \eqref{eq:I16} has been dealt with extensively for a particular value of $M$ \cite{Wolfram_meijer}, a general solution is not available in the literature. The challenge here is to deal with the Meijer-G function with a positive integer.
	In the following Proposition \ref{prop1}, we introduce a novel approach to solve \eqref{eq:I16}:
	\begin{my_proposition}\label{prop1}
		We represent the solution of integral involving  Meijer-G function with a positive integer in \eqref{eq:I16}  using multivariate Fox-H function:
		\begin{flalign}\label{eq:I12}
			&			I =   \phi_1(M) H_{p_0,q_0:(p,q)_{i=1}^M}^{0,n_0:(m,n)_{i=1}^M}  \left( \begin{array}{c} \{(a_{w},A_{w})_{w=1}^{p}\}_{i=1}^{M}\\ \{(b_{w},B_{w})_{w=1}^{q}\}_{i=1}^{M} \end{array}  \middle\vert \{\psi x\}_{i=1}^M \right) 
		\end{flalign}
		where $\phi_1(M)$ is a function.
	\end{my_proposition}       
	\begin{IEEEproof}
		Using the Meijer-G definition of \eqref{eq:mg_def} in \eqref{eq:I16}, we get
		\begin{flalign}\label{eq:I15}
			&I=\int_{0}^{\infty} f(x) \left(\frac{1}{2\pi j}\int_L F(s) (\psi x)^s ds\right)^{M}  dx 
		\end{flalign}
		A caution is not to apply Fubini's theorem directly to interchange the integrals in \eqref{eq:I15} to solve the inner integral in $x$. First, we need to represent \eqref{eq:I15} as the product of   $M$ contour integrals
		\begin{flalign}\label{eq:I116}
			&I= \int_{0}^{\infty} f(x) \bigg[\frac{1}{(2\pi j)^N} \oint \limits_{L}  F(s)\psi^{s} ds \cdots \oint \limits_{L} F(s)\psi^{s} ds \bigg]  dx 
		\end{flalign}
		Next, we apply the theory of multiple integrals and Fubini's theorem to interchange the integral to express  \eqref{eq:I116}
		\begin{flalign}\label{eq:I201}
			&I=\frac{1}{(2\pi j)^M} \oint \limits_{L} \cdots \oint \limits_{L} \bigg[\int_{0}^{\infty} f(x) x^{\sum_{i}^M s_i} dx   \nonumber \\ & \times F(s_1)\psi^{s_1} \cdots F(s_M)\psi^{s_M} ds_1\cdots ds_M\bigg]
		\end{flalign}
		Finally, solving the inner integral  as $\int_{0}^{\infty} \phi(x) x^{\sum s_i} dx= \phi(M)G(\sum s_i)$ in terms of Gamma function, we can get \eqref{eq:I12}.
		
	\end{IEEEproof}
	The representation in \eqref{eq:I12} is a particular type of a general multivariate Fox-H function: it involves product $M$-Mellin-Barnes integrals, each with the same integrand over the same contour except with a single additional linear combination term.
	
	To circumvent the linear combination term, we use a condition $s_1=s_2=s_3=\cdots=s_M$ to solve \eqref{eq:I12} using a single-variate Fox-H function. The assumption  $s_1=s_2=s_3=\cdots=s_M$ is reasonable for the i.i.d. model providing a computationally efficient single-variate representation for  performance evaluation  of the UOWC system \cite{blackseacom_2024}. 
	\begin{my_proposition}\label{prop3}
		Assuming $s_1=s_2=s_3=\cdots=s_M$, we can solve \eqref{eq:I16}  using a single-variate Fox-H function as
		\begin{flalign}\label{eq:I121}
			&	I
			\approx \phi_2(M) \left(G_{p,q}^{m,n}  \left(\begin{array}{c} \{a_{w}\}_{w=1}^{p}\\ \{b_{w}\}_{w=1}^{q} \end{array} \middle\vert \psi  \right)\right)^{M-1}  \nonumber \\& \times  H_{p',q'}^{m',n'}  \left(\begin{array}{c} \{(a'_{w},A_{w})\}_{w=1}^{p'}\\ \{(b'_{w},B_{w})\}_{w=1}^{q'}  \end{array} \middle\vert \psi  \right)
		\end{flalign}
		where $\phi_2(M)$  is a function.
	\end{my_proposition}
	\begin{IEEEproof}
		We represent \eqref{eq:I16} as
		\begin{flalign}\label{eq:I151}
			&I=\int_{0}^{\infty} f(x) \left(G_{p,q}^{m,n}  \left(\begin{array}{c} \{a_{w}\}_{w=1}^{p}\\ \{b_{w}\}_{w=1}^{q} \end{array} \middle\vert \psi x \right)\right)^{M-1} \nonumber \\& \times  \left(G_{p,q}^{m,n}  \left(\begin{array}{c} \{a_{w}\}_{w=1}^{p}\\ \{b_{w}\}_{w=1}^{q} \end{array} \middle\vert \psi x \right)\right) dx 
		\end{flalign}
		Using the Meijer-G definition of \eqref{eq:mg_def} in \eqref{eq:I151}, we get
		
		\begin{flalign}\label{eq:I201_apx}
			&I=\int_{0}^{\infty} f(x) \left(\frac{1}{2\pi j}\int_L F(s) (\psi x)^s ds\right)^{M-1} \nonumber \\& \times  \left(\frac{1}{2\pi j}\int_L F(s) (\psi x)^s ds\right) dx 
		\end{flalign}
		Assuming $s_1=s_2=s_3=\cdots=s_M$, and 	applying the Fubini's theorem, we express \eqref{eq:I201_apx}:
		\begin{flalign}\label{eq:I2011}
			&I\approx \left(\frac{1}{2\pi j}\int_L F(s) (\psi)^s ds\right)^{M-1}  \nonumber \\& \times  \left(\frac{1}{2\pi j}\int_L \bigg[\int_{0}^{\infty} x^{Ms}\phi(x) dx\bigg] F(s)\psi^s ds\right) 
		\end{flalign}
		Solving the inner integral $\int_{0}^{\infty} x^{Ms}\phi(x) dx= \phi_2(M)G(s)$ in terms of Gamma function,  we get \eqref{eq:I121}.
	\end{IEEEproof}
	Numerical evaluation conducted in Section VI validates the accuracy of approximate analysis as presented in \eqref{eq:I121}.
	
	\section{Performance Analysis}
	In this section, we use the general approach developed in Section III  to derive exact analytical expressions for average BER and ergodic capacity for SC-based multi-aperture UOWC systems for both i.ni.d. and i.i.d. channels. First, we focus on the average BER performance with exact, approximate, and asymptotic analytical results in terms of system parameters for two scenarios of underwater turbulence parameterized by $\omega$. Next, we derive an analytical expression for the ergodic capacity.

	\subsection{Average BER}
	A general formulation for the average BER over fading channels  using the CDF $F_{\gamma_{\rm SC}}(\gamma)$ is given by \cite{Ansari2011_ber} 
	\begin{eqnarray}\label{BER1}
		\bar{P}_{e} = \frac{q^p}{2\Gamma(p)} \int_{0}^{\infty} e^{-q\gamma} \gamma^{p-1} F_{\gamma}(\gamma)  d\gamma
	\end{eqnarray}
	where $p$ and $q$ define different modulation schemes.
	
	\begin{my_lemma}
		An analytical expression for 	average BER for SC-based UOWC with i.ni.d. channels is given by
		\begin{flalign}\label{cdf_inid_pe10}
			&\bar{P}_{e} = \frac{\rho^{2N}}{2\Gamma(p)}
			\sum_{S \subseteq \{1,..,N\}} \prod_{i \epsilon S, i\neq j,k} \omega_i  \prod_{j \epsilon S^C, j \neq i,k} \frac{1-\omega_j}{c_j\Gamma(a_j)} \nonumber \\ &\times	H^{(0,1):(2,1)_1;...(2,1)_N}_{(1,0):(2,3)_1;...(2,3)_N} \left[ \begin{array}{c} V_1 \\ \vdots \\ V_N \end{array} \middle \vert \begin{array}{c} U_1 \\ U_2 \end{array} \right]
		\end{flalign}
		where $U_1 = \{(1-p; \{\alpha_{l}\}_{l = 1}^N ): ((c^1_l,\gamma^1_l), (c^2_l,\gamma^2_l))_{l = 1}^N \}$ and $U_2 = \{(-; -): ((d^1_l,\delta^1_l),..., (d^3_l,\delta^3_l))_{l = 1}^N\} $. If $l \in S $, then $ \alpha_l = \frac{1}{2}$ $((c^1_l,\gamma^1_l),(c^2_l,\gamma^2_l)) = ((1, 1),(\rho_l^2 +1, 1))$, $((d^1_l,\delta^1_l),..., (d^3_l,\delta^3_l)) = ((1,1), (\rho_l^2,1), (0,1))$, and $V_l = \psi_l q^{c_l/2} $. If $l \in S^c $, then
		$\alpha_l = \frac{c_l}{2} $, $ ((c^1_l,\gamma^1_l),(c^2_l,\gamma^2_l)) = ((1, 1),(\frac{\rho_l^2}{c_l} +1, 1)) $ and $((d^1_l,\delta^1_l),..., (d^3_l,\delta^3_l)) = ((a_l,1), (\frac{\rho_l^2}{c_l},1), (0,1)) $ and $V_l = \psi_l q^{c_l/2}$.
	\end{my_lemma}
	\begin{IEEEproof}
		The proof is presented in Appendix A.
	\end{IEEEproof}
Using the results from \cite{Abo2018}, the average BER performance at high SNR $\bar{\gamma}\to \infty$ can be expressed asymptotically as $\bar{P}e^{\infty}= G_c {\bar{\gamma}}^{-G_d}$, where $G_c$ represents the coding gain (which can be represented using system parameters) and $G_d$ is the diversity order: 
\begin{flalign}\label{eq:inid_do}
	 G_d= \sum_{i=1}^N\min\{\frac{1}{2}, \frac{a_ic_i}{2}, \frac{\rho_i^2}{2}\}. 
\end{flalign} 
The diversity order indicates that the performance of the multi-aperture UOWC system can be enhanced by increasing the number of apertures. Also, pointing errors can be mitigated for a given turbulence level by adjusting the beam-width to achieve a higher value of $\rho^2$.
	
The multivariate Fox-H function has gained popularity for analyzing the statistical performance of wireless systems under complex fading models. This representation is favored for two reasons: it offers an exact solution where conventional mathematical methods do not yield the desired results, and its asymptotic results provide valuable analytical insights into system performance and diversity order.

	\begin{my_lemma}
		An analytical expression for average BER for SC-based UOWC with i.i.d. channels is given by
		\begin{flalign}\label{ber_iid_final}
			&\bar{P}_e = \frac{q^p \rho^2}{2\Gamma(p)} \sum\limits_{k = 0}^{N}\binom{N}{k} \omega^{N-k} \left(\frac{1-\omega}{c \Gamma(a)}\right)^{k} \nonumber \\ & \times  H_{(1,0):(2,3)_1;... (2,3)_N}^{(0,1):(2,1)_1;... (2,1)_N} \left[ \begin{array}{c} U_1 \\ U_2 \end{array} \middle \vert \begin{array}{c}V_1 \\ \vdots \\ V_N \end{array} \right]   
		\end{flalign}
		where $U_1 = \left(1-p; \left(\frac{1}{2}\right)_{1,k};\left(\frac{c}{2}\right)_{k+1,N}\right):{((a_1 \alpha_1) (a_2, \alpha_2))_i}_{i =1}^N $ and $U_2 =  - : {((b_1,\beta_1),(b_2,\beta_2),(b_3,\beta_3))_i}_{i = 1}^N$. 
		If  $i \leq N-k$ then  $((a_1,\alpha_1)(a_2, \alpha_2))_i = ((1,1)(\rho^2+1, 1)) $, $((b_1,\beta_1),(b_2,\beta_2),(b_3,\beta_3))_i = ((1,1),(\rho^2,1),(0,1))$, and $V_i = \psi_1 q^{1/2}$. If  $i \geq N-k$ then $((a_1,\alpha_1)(a_2, \alpha_2))_i = ((1,1)(\frac{\rho^2}{c}+1, 1)) $, $((b_1,\beta_1),(b_2,\beta_2),(b_3,\beta_3))_i = ((a,1),(\frac{\rho^2}{c},1),(0,1))$, and $V_i = \psi_2 q^{c/2}$.
	\end{my_lemma}
	\begin{IEEEproof}
		Using \eqref{cdf_uw_pe} and \eqref{cdf_sc_iid} in \eqref{BER1}, and applying the binomial expansion, we get
		\begin{flalign}\label{cdf_uw_pe2}
			&\bar{P}_e = \frac{q^p \rho^2}{2\Gamma(p)} \sum\limits_{k = 0}^{N}\binom{N}{k} \omega^{N-k} \left(\frac{1-\omega}{c \Gamma(a)}\right)^{k} \int_{0}^{\infty} e^{-q\gamma} \gamma^{p-1} \nonumber \\ & \times \bigg( G_{2,3}^{2,1}\left(\begin{array}{c}1,\rho^2+1\\1,\rho^2,0\end{array}\left|\frac{1}{\lambda A} \left(\sqrt{\frac{\gamma}{\bar{\gamma}}}\right)\right.\right) \bigg)^{N-k} \nonumber \\ & \times \bigg (G_{2,3}^{2,1}\left(\begin{array}{c}1,\frac{\rho^2}{c}+1\\a,\frac{\rho^2}{c},0\end{array}\left|\frac{1}{b^{c} A^{c}}\left(\sqrt{\frac{\gamma}{\bar{\gamma}}}\right)^{c}\right.\right) \bigg)^{k}  d\gamma
		\end{flalign}
		Applying the result of the Meijer-G function with integer exponent  of Proposition \ref{prop1},  \eqref{cdf_uw_pe2} becomes:
		\begin{flalign}\label{cdf_uw_pe4}
			&\bar{P}_e = \frac{q^p \rho^2}{2\Gamma(p)} \sum\limits_{k = 0}^{N}\binom{N}{k} \omega^{N-k} \left(\frac{1-\omega}{c \Gamma(a)}\right)^{k} \oint\limits_{L_1} \cdots \oint\limits_{L_N}  \int_{0}^{\infty} e^{-q\gamma}  \nonumber \\ & \times \gamma^{p-1+(s_1+\cdots +s_k+c(s_{k+1}..+s_N))/2} d\gamma F(s_1) \psi_1 ^{s_1} \cdots F(s_{N-k}) \nonumber \\ & \times  \psi_1^{s_{N-k}}F(s_{N-k+1}) \psi_2^{s_{N-k+1}} \cdots F(s_{N}) \psi_2^{s_{N}}   ds_1\cdots ds_N
		\end{flalign}
		where	$\psi_1 = \frac{1}{\lambda A} \left(\sqrt{\frac{1}{\bar{\gamma}}}\right)$ and  $\psi_2 = \frac{1}{b^{c} A^{c}}\left(\sqrt{\frac{1}{\bar{\gamma}}}\right)^{c}$. Solving the inner integral in terms of the Gamma function with the application of multivariate Fox-H function representation,  we get  \eqref{ber_iid_final}.
	\end{IEEEproof}
Applying the results from \cite{Abo2018}, we can represent the average BER performance at  high SNR $\bar{\gamma}\to \infty$ as  $\bar{P}_e^{\infty}= G_c {\bar{\gamma}}^{-G_d}$ asymptotically, where $G_c$ is the coding gain and $G_d$ is the diversity order	$G_d= \min\{\frac{N}{2}, \frac{Nac}{2}, \frac{N\rho^2}{2}\}$.

	Although the multivariate Fox-H function can be computed using MATLAB/PYHON routines, it is desirable to provide analysis using the single-variate Fox-H function. In the following Lemma, we approximate the average BER performance for i.i.d. channels using the single-variate Fox-H function.
	
	\begin{my_lemma}
		An approximate expression for the average BER for SC-based UOWC with i.i.d. channels is given by
		\begin{flalign}\label{eq:lemma_final}
			\bar{P}_{e}\approx &\frac{q\rho^{2N}}{2q^{p}\Gamma(p)}\sum\limits_{k = 0}^{N}\binom{N}{k}\omega^{N-k}\big(\frac{(1-\omega)}{c \Gamma(a)}\big)^k \nonumber \\ & \times \Bigg(G_{2,3}^{2,1}\left(\begin{array}{c}1,\rho^2+1\\1,\rho^2,0\end{array}\left|\frac{1}{\lambda A}
			\left(\sqrt{\frac{1}{\bar{\gamma}}}\right)\right.\right)\Bigg)^{N-k-1} \nonumber \\ & \times \Bigg( G_{2,3}^{2,1}\left(\begin{array}{c}1,\frac{\rho^2}{c}+1\\a,\frac{\rho^2}{c},0\end{array}\left| \frac{1}{b^{c} A^{c}}\left(\sqrt{\frac{1}{\bar{\gamma}}}\right)^{c} \right.\right)\Bigg)^{k-1}\nonumber \\& \times H_{1,0:2,3;2,3}^{0,1:2,1;2,1}\left[\begin{array}{c} V_1\\V_2\end{array}\bigg| \begin{array}{c} \frac{1}{\lambda A \sqrt{q^{N-k}\gamma_{0}}} \\ \frac{1}{b^{c} A^{c}}\left(\sqrt{\frac{1}{q^{k}\bar{\gamma}}}\right)^{c}\end{array}\right]
		\end{flalign}
		where $V_1=(1-p:\frac{N-k}{2},\frac{kc}{2}):(1,1),(1,\rho^2+1);(1,1),(1,\frac{\rho^2}{c}+1)$ and $V_2=-:(1,1),(1,\rho^2),(1,0);(1,a),(1,\frac{\rho^2}{c}),(1,0)$.
	\end{my_lemma}
	\begin{IEEEproof}
		Using \eqref{cdf_uw_pe} in  \eqref{cdf_sc_iid} and substituting the resulting expression in  \eqref{BER1} with the application of binomial expansion, we get
		\begin{eqnarray}\label{ber1}
			&\bar{P}_{e}=\frac{q\rho^{2N}}{2\Gamma(p)}\sum\limits_{k = 0}^{N}\binom{N}{k}\omega^{N-k}\big(\frac{(1-\omega)}{c \Gamma(a)}\big)^k \nonumber \\& \times \int_{0}^{\infty}e^{-q\gamma} \gamma^{p-1}  \Bigg(G_{2,3}^{2,1}\left(\begin{array}{c}1,\rho^2+1\\1,\rho^2,0\end{array}\left|\frac{1}{\lambda A}
			\left(\sqrt{\frac{\gamma}{\bar{\gamma}}}\right)\right.\right)\Bigg)^{N-k}\nonumber\\& \times \Bigg( G_{2,3}^{2,1}\left(\begin{array}{c}1,\frac{\rho^2}{c}+1\\a,\frac{\rho^2}{c},0\end{array}\left|\frac{1}{b^{c} A^{c}}\left(\sqrt{\frac{\gamma}{\bar{\gamma}}}\right)^{c}\right.\right)\Bigg)^{k}d\gamma
		\end{eqnarray}
		Applying the analytical approach of Proposition 	\ref{prop3} in \eqref{ber1} yields \eqref{eq:lemma_final}, which completes the proof.
	\end{IEEEproof}
	The bivariate Fox-H function has computational routines available in literature \cite{Elmehdi2017}. In order to provide an insight into the high SNR regime, we use the series expansion of the bivariate Fox-H function to get 
	\begin{flalign}\label{eq:assymp_gen}
		&\bar{P}_{e}^{
			\infty}=\frac{q\rho^{2N}}{2q^{p}\Gamma(p)}\sum\limits_{k = 0}^{N}\binom{N}{k}\omega^{N-k}\big(\frac{(1-\omega)}{c \Gamma(a)}\big)^k \nonumber \\ & \times \Bigg(\frac{\Gamma(1-p_{1})\Gamma(\rho^2-p_{1})\Gamma(p_{1})}{\Gamma(1+\rho^2-p_{1})\Gamma(1+p_{1})} \bigg(\frac{1}{\lambda A \sqrt{q^{N-k}\gamma_{0}}}\bigg)^{p_{1}}\Bigg)^{N-k} \nonumber \\ & \times\Bigg( \frac{\Gamma(a-p_{2})\Gamma(\frac{\rho^2}{c}-p_{2})\Gamma(p_{2})}{\Gamma(1+\frac{\rho^2}{c}-p_{2})\Gamma(1+p_{2})} \bigg(\frac{1}{b^{c} A^{c}}\left(\sqrt{\frac{1}{q^{k}\bar{\gamma}}}\right)^{c}\bigg)^{p_{2}}\Bigg)^{k}\nonumber \\& \times
		\Gamma(\frac{N-k}{2}p_1+\frac{kc}{2}p_2+p)
	\end{flalign}
	where $p_{1}=\min\{1, \rho^2\}$ and $p_{2}=\min\{ac, \rho^2\}$. Using \eqref{eq:assymp_gen}, we can get the diversity order of the system as $\text{DO}^{}=\min\{\frac{N}{2}, \frac{Nac}{2}, \frac{N\rho^2}{2}\}$. The diversity order demonstrates that the average BER performance can be improved by increasing the aperture at the detector of the UOWC system.
	
	The parameter $\omega$ in the EEG model signifies the mixture of exponential and generalized gamma distributions. Lower $\omega$ denotes the lower bubble level in the oceanic channel. Further, $\omega=0$ in \eqref{cdf_uw_pe} reduces the CDF function to a single Meijer-G, a typical scenario for terrestrial fading channels. Thus, applying the straightforward application of  Proposition \eqref{prop3}, we get 
	
	\begin{flalign}\label{eq:lemma2_final}
		\bar{P}_{e}\approx& \frac{1}{2 \Gamma(p)}\big(\frac{\rho^2}{c \Gamma(a)}\big)^N \nonumber \\ & \times \Bigg(G_{2,3}^{2,1}\left(\begin{array}{c}1,\frac{\rho^2}{c}+1\\a,\frac{\rho^2}{c},0\end{array}\left|\frac{1}{b^{c} A^{c}}\left(\sqrt{\frac{1}{\bar{\gamma}}}\right)^{c}\right.\right)\Bigg)^{N-1}\nonumber \\ & \times
		H_{3,3}^{2,2}\left(\begin{array}{c}V_{3}\\V_{4}\end{array}\left|\frac{1}{b^{c} A^{c}}\left(\sqrt{\frac{1}{q^{N}\bar{\gamma}}}\right)^{c}\right.\right)
	\end{flalign}
	where $V_{3}=(1,1),(1-p,\frac{Nc}{2}),(\frac{\rho^2}{c}+1, 1)$ and $V_{4}=(a,1),(\frac{\rho^2}{c}, 1),(0, 1)$.
	Further, an asymptotic expression for the average BER with $\omega=0$ can also be derived using the series expansion of the Meijer-G function at $\bar{\gamma} \to \infty$:
	\begin{flalign}\label{eq:assymp_gen2}
		&\bar{P}_{e}^{
			\infty}=\big(\frac{q^{-p}\rho^2}{c \Gamma(a)}\big)^N \Bigg(  \bigg(\frac{1}{b^{c} A^{c}}\left(\sqrt{\frac{1}{q^{N}\bar{\gamma}}}\right)^{c}\bigg)^{p_{3}}\Bigg)^{N}\nonumber \\ & \times \Bigg(\frac{\Gamma(a-p_{3})\Gamma(\frac{\rho^2}{c}-p_{3})\Gamma(p_{3})\Gamma(p+\frac{Nc}{2}p_{3})}{\Gamma(\frac{\rho^2}{c}+1-p_{3})\Gamma(1+p_{3})}\Bigg)^{N}
	\end{flalign}
	where $p_{3}=\min\{a, \frac{\rho^{2}}{c}\}$.
	\subsection{Ergodic Capacity}
	The ergodic capacity $C$ is the  expected value of channel capacity $\mathbb{E}[\log_2(1+\gamma)]$, where $\mathbb{E}$ is the expectation operator:
	\begin{eqnarray} \label{eq:cap_eq} 
		C = \int_{0}^{\infty} \log_2 (1+\gamma) f_{\gamma} (\gamma) d\gamma
	\end{eqnarray}
	The analysis for the ergodic capacity becomes more challenging since the PDF for the i.ni.d. model becomes after  substituting \eqref{pdf_uw_pe}  in \eqref{pdf_sc_inid}:
	\begin{flalign}\label{eq:cap_inid_f1_zaf}
		f_{\gamma} (\gamma) &= \sum_{k =1}^{N} \frac{\omega_k \rho_k^{2n} }{2 \gamma} \prod_{n \neq k} \left(F_{X_n}(x)\right)  \nonumber \\& \times G_{1,2}^{2,0}\left(\begin{array}{c}\rho_k^2+1\\1,\rho_k^2\end{array}\left|\frac{1}{\lambda_k A_k} \left(\sqrt{\frac{\gamma}{\bar{\gamma}}}\right)\right.\right) \nonumber \\& +\frac{(1-\omega_k)\rho_k^{2n}}{2 \Gamma(a_k) \gamma} \prod_{n \neq k} \left(F_{X_n}(x)\right) \nonumber \\&  \times G_{1,2}^{2,0}\left(\begin{array}{c}\frac{\rho_k^2}{c_k}+1\\a_k,\frac{\rho_k^2}{c_k}\end{array}\left|\frac{1}{b_k^{c_k} A_k^{c_k}}\left(\sqrt{\frac{\gamma}{\bar{\gamma}_k}}\right)^{c_k}\right.\right)  
	\end{flalign}
	where  $F_{X_n}(x)$ is defined as in \eqref{cdf_uw_pe}.
	In the following Lemma, we present an analytical expression for the ergodic capacity by representing \eqref{eq:cap_eq} with  \eqref{eq:cap_inid_f1_zaf} and \eqref{eq:cap_inid_f3} in terms of multivariate Fox-H function.
	\begin{my_lemma}
		An analytical expression for the ergodic capacity  of the SC-based UOWC system  with i.ni.d. channels is given by \eqref{eq:cap_inid_exact}
		\begin{figure*}
			\begin{flalign} \label{eq:cap_inid_exact}
				C =& \sum_{k =1}^{N} \frac{\omega_k \prod_{n = 1}^N (\rho_n^{2}) }{\ln(2)} \lambda_k^2A_k^2\bar{\gamma}_{k} \sum_{S} \prod_{i \epsilon S, i\neq j,k} \omega_i  \prod_{j \epsilon S^C, j \neq i,k} \frac{1-\omega_j}{c_j\Gamma(a_j)}    H^{(0,2):(1,2);(2,1)_1;...(2,1)_{N-1}}_{(2,1):(2,2);(2,3)_1;...(2,3)_{N-1}} \nonumber \\ &\left[ \begin{array}{c}V_0 \\ \{ V_l\}_{l=1}^{N-1} \end{array} \middle \vert \begin{array}{c} \left(-2;2,\left \{  {\alpha}_1^{l} \right \}_{l=1}^{N-1} \right), \left(-\rho_k^2-1; 2,  \left \{  {\alpha}_2^{l} \right \}_{l=1}^{N-1}\right): ((0,1),(0,1));\left\{((c^1_l,\gamma^1_l), (c^2_l,\gamma^2_l))\right\}_{l=1}^{N-1}  \\ \left(-\rho_k^2-2; 2, \left \{  {\beta}_1^{l} \right \}_{l=1}^{N-1} \right):((0,1),(-1,1)); \left\{ ((d^1_l,\delta^1_l),(d^2_l,\delta^2_l), (d^3_l,\delta^3_l)) \right\}_{l=1}^{N-1}  \end{array}\right] \nonumber \\ & +	\frac{(1-\omega_k) \prod_{n = 1}^N (\rho_n^{2}) }{2 c_k \Gamma(a_k)ln(2)} b_k^2A_k^2\bar{\gamma}_{k} \sum_{S} \prod_{i \epsilon S, i\neq j,k} \omega_i  \prod_{j \epsilon S^C, j \neq i,k} \frac{1-\omega_j}{c_j\Gamma(a_j)} 
				H^{(0,2):(1,2);(2,1)_1;...(2,1)_{N-1}}_{(2,1):(2,2);(2,3)_1;...(2,3)_{N-1}}  \nonumber \\ & \left[ \begin{array}{c}V_0' \\\{ {V'}_l\}_{l=1}^{N-1} \end{array} \middle \vert \begin{array}{c} \left(1-a_k-\frac{2}{c_k};\frac{2}{c_k}, \left \{  {\alpha'}_1^{l} \right \}_{l=1}^{N-1} \right), \left(1-\frac{\rho_k^2+2}{c_k}; 2,  \left \{  {\alpha'}_2^{l} \right \}_{l=1}^{N-1} \right): ((0,1),(0,1));\left\{((c^1_l,\gamma^1_l), (c^2_l,\gamma^2_l)) \right\}_{l=1}^{N-1} \\ \left(-\frac{\rho_k^2+2}{c_k}; \frac{2}{c_k},\left \{  {\beta'}_1^{l} \right \}_{l=1}^{N-1} \right):((0,1),(-1,1)); \left\{ ((d^1_l,\delta^1_l),(d^2_l,\delta^2_l), (d^3_l,\delta^3_l)) \right\}_{l=1}^{N-1}  \end{array}\right]
			\end{flalign}
			
			If $l \in S$, then ${\alpha}_1^{l} = {\alpha}_2^{l}={\alpha'}_1^{l} = {\alpha'}_2^{l} = {\beta}_1^{l} = {\beta}_1^{l} = 1 $, $ ((c^1_l,\gamma^1_l), (c^2_l,\gamma^2_l)) = \left((1,1), \left(\frac{\rho^2_i}{c_i},1\right)\right) $, $ ((d^1_l,\delta^1_l),(d^2_l,\delta^2_l), (d^3_l,\delta^3_l)) = \left((a_i, 1), \left( \frac{\rho_i^2}{c_i}, 1\right)\right) $ and $V_l = \frac{\lambda_kA_k}{\lambda_i A_i} \sqrt{\frac{\bar{\gamma}_k}{\bar{\gamma}_i}}$ , $ {V'}_l = \left(\frac{\lambda_k A_k}{b_j A_j} \sqrt{\frac{\bar{\gamma}_k}{\bar{\gamma}_j}}\right)^{c_j}$.
			
			If $l \in S^c$, then ${\alpha}_1^{l} = {\alpha}_2^{l}={\alpha'}_1^{l} = {\alpha'}_2^{l} = {\beta}_1^{l} = {\beta}_1^{l} = c_j $, $ ((c^1_l,\gamma^1_l), (c^2_l,\gamma^2_l)) = \left((1,1), \left(\frac{\rho^2_i}{c_i},1\right)\right) $, $ ((d^1_l,\delta^1_l),(d^2_l,\delta^2_l), (d^3_l,\delta^3_l)) = \left((a_i, 1), \left( \frac{\rho_i^2}{c_i}, 1\right)\right) $ and $V_l = \frac{b_kA_k}{\lambda_i A_i} \sqrt{\frac{\bar{\gamma}_k}{\bar{\gamma}_i}}$ , $ {V'}_l = \left(\frac{b_k A_k}{b_j A_j} \sqrt{\frac{\bar{\gamma}_k}{\bar{\gamma}_j}}\right)^{c_j}$. 
			
			\hrule
		\end{figure*}
	\end{my_lemma} 
	\begin{IEEEproof}
		The proof is presented in Appendix B.
	\end{IEEEproof}
	Next, we analyze the ergodic capacity for the i.i.d. model. Since the analysis approach will be similar, we consider the case with  $\omega = 0$. Using \eqref{pdf_uw_pe},  \eqref{cdf_uw_pe}, and \eqref{pdf_sc_iid} in  \eqref{eq:cap_eq}, we get 
	\begin{flalign} \label{eq:cap_iid_2} 
		& C = \frac{Nc}{2} \left(\frac{ \rho^{2}}{c\Gamma(a)}\right)^N  \int_{0}^{\infty} \frac{ \log_2 (1+\gamma)}{\gamma} \nonumber \\ & \times \bigg[ G_{2,3}^{2,1}\left(\begin{array}{c}1,\frac{\rho^2}{c}+1\\a,\frac{\rho^2}{c},0\end{array}\left|\frac{1}{b^{c} A^{c}}\left(\sqrt{\frac{\gamma}{\bar{\gamma}}}\right)^{c}\right.\right) \bigg]^{N-1} \nonumber \\ & \times G_{1,2}^{2,0}\left(\begin{array}{c}\frac{\rho^2}{c}+1\\a,\frac{\rho^2}{c}\end{array}\left|\frac{1}{b^{c} A^{c}}\left(\sqrt{\frac{\gamma}{\bar{\gamma}}}\right)^{c}\right.\right) d\gamma
	\end{flalign}
	Conventional approach uses $\log_2(1+\gamma)$ in the inner integral to solve \eqref{eq:cap_iid_2}. However, the resultant inner integral cannot be represented in terms of the Gamma function for a compatible definition of the multivariate Fox-H function. We approach it differently to get a closed-form expression of the ergodic capacity for the i.i.d. channel model, as given in the following Lemma.
	\begin{my_lemma}
		An analytical expression for the ergodic capacity  of the SC-based UOWC system  with i.i.d. channels and $\omega=0$ is given by
		\begin{flalign}\label{eq:cap_iid_8}
			& C =    \frac{Nb^2A^2\gamma_{0}} {ln(2)} \left(\frac{\omega\rho^2}{c \Gamma(a)}\right)^N H^{0,2:1,2;(2,1)_{1, N-1}}_{2,1:2,2;(2,3)_{1, N-1}}   \left[ \begin{array}{c}b^2A^2\gamma_{0} \\ \{1\}_{N-1} \end{array} \middle \vert \begin{array}{c} V_1 \\ V_2 \end{array}\right] 
		\end{flalign}
		where $V_1 = (1-a-\frac{2}{c};\frac{2}{c}, 1_{1, N-1} ), (1-\frac{\rho^2}{c} - \frac{2}{c}; \frac{2}{c}, 1_{1, N-1}): (0,1), (0,1); (1,1), (\frac{\rho^2}{c}+1, 1) $ and $V_2 =  (-\frac{\rho^2}{c}-\frac{2}{c}; \frac{2}{c}, 1_{1, N-1}): (0, 1), (-1,1); (a,1), (\frac{\rho^2}{c}, 1), (0,1) $
	\end{my_lemma}
	\begin{IEEEproof}
		The proof is presented in Appendix C.
	\end{IEEEproof}
	
	Finally, we use the Proposition \ref{prop3} to represent the ergodic capacity using a single-variate Fox-H function
	\begin{my_lemma}
		An approximate expression for the ergodic capacity  of the SC-based UOWC system  with i.i.d. channels and $\omega=0$ is given by
		\begin{flalign}\label{cap_final_iid}
			& C \approx  \frac{Nb^2A^2\gamma_{0}} {ln(2)} \left(\frac{\omega\rho^2}{c \Gamma(a)}\right)^N   H^{0,2:1,2;2,1}_{2,1:2,2;2,3} \bigg[ \begin{array}{c}b^2A^2\gamma_{0} \\ 1 \end{array} \bigg \vert  \begin{array}{c} V_1 \\ V_2  \end{array}\bigg]  \nonumber \\ & \times \bigg[G_{2,3}^{2,1}\left(\begin{array}{c}1,\frac{\rho^2}{c}+1\\a,\frac{\rho^2}{c},0\end{array}\left|\frac{1}{b^{c} A^{c}}\left(\sqrt{\frac{\gamma}{\bar{\gamma}}}\right)^{c}\right.\right) \bigg ]^{N-2} 
		\end{flalign}
		where $V_1=(1-a-\frac{2}{c};\frac{2}{c}, N-1 ), (1-\frac{\rho^2}{c} - \frac{2}{c}; \frac{2}{c}, N-1): (0,1), (0,1); (1,1), (\frac{\rho^2}{c}+1, 1)$ and $V_2=(-\frac{\rho^2}{c}-\frac{2}{c}; \frac{2}{c}, N-1): (0, 1), (-1,1); (a,1), (\frac{\rho^2}{c}, 1), (0,1)$.
	\end{my_lemma}
	\begin{IEEEproof}
		The proof is presented in Appendix D.
	\end{IEEEproof}

	\begin{figure}[h!]
		\centering
		\includegraphics[scale=0.395]{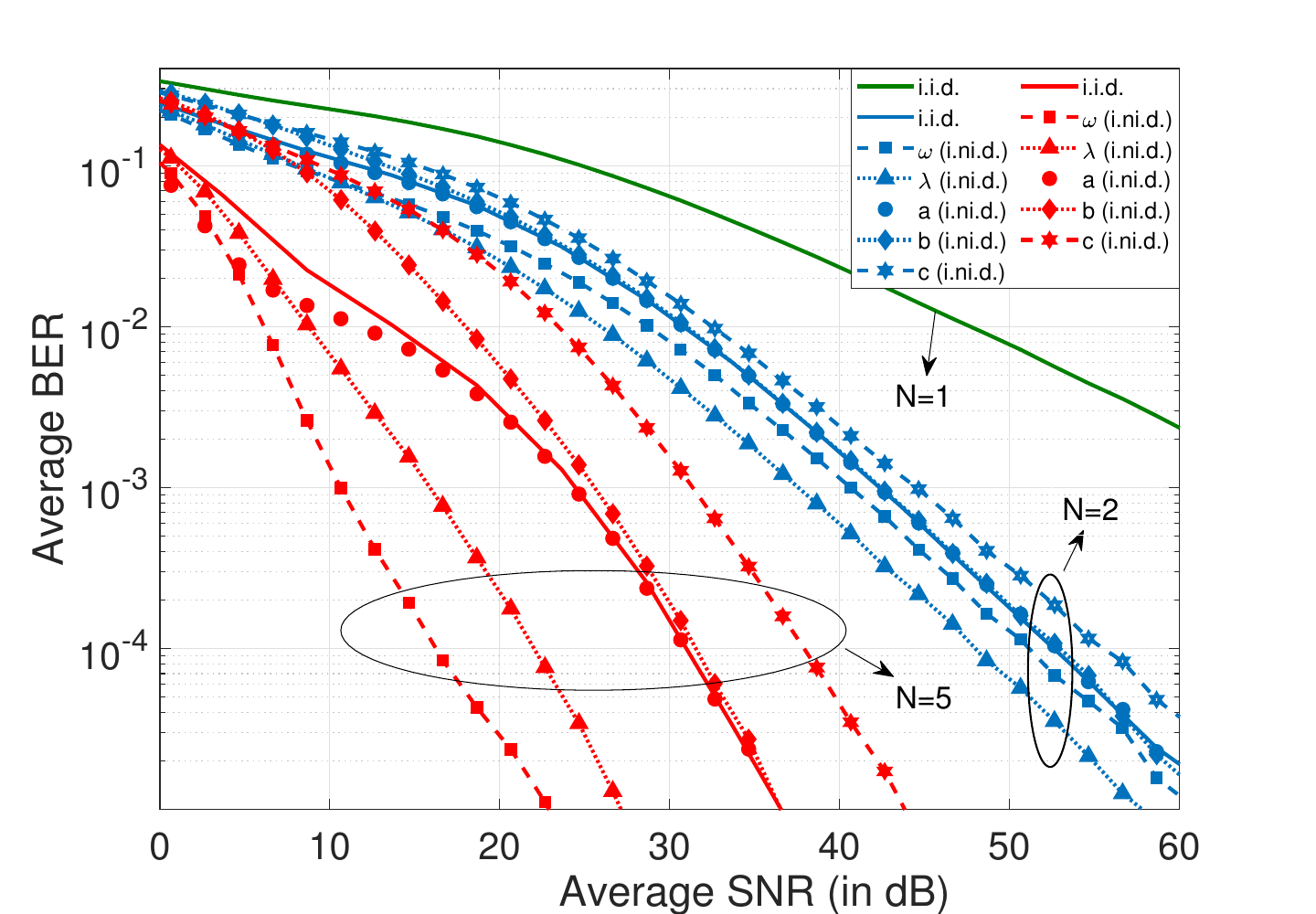}
		\caption{Average BER performance for i.ni.d. UOWC system with pointing errors by adopting distinct values of the EGG channel for each link with $N=1$, $N=2$, and  $N=5$. The pointing error parameters are $A_0=0.1639$ and $\rho=0.9875$. Simulation results are close to the analytical numerical results (not shown for brevity). }
		\label{fig:aber_inid}
	\end{figure}

	\begin{figure}[h!]
		\includegraphics[scale=0.395]{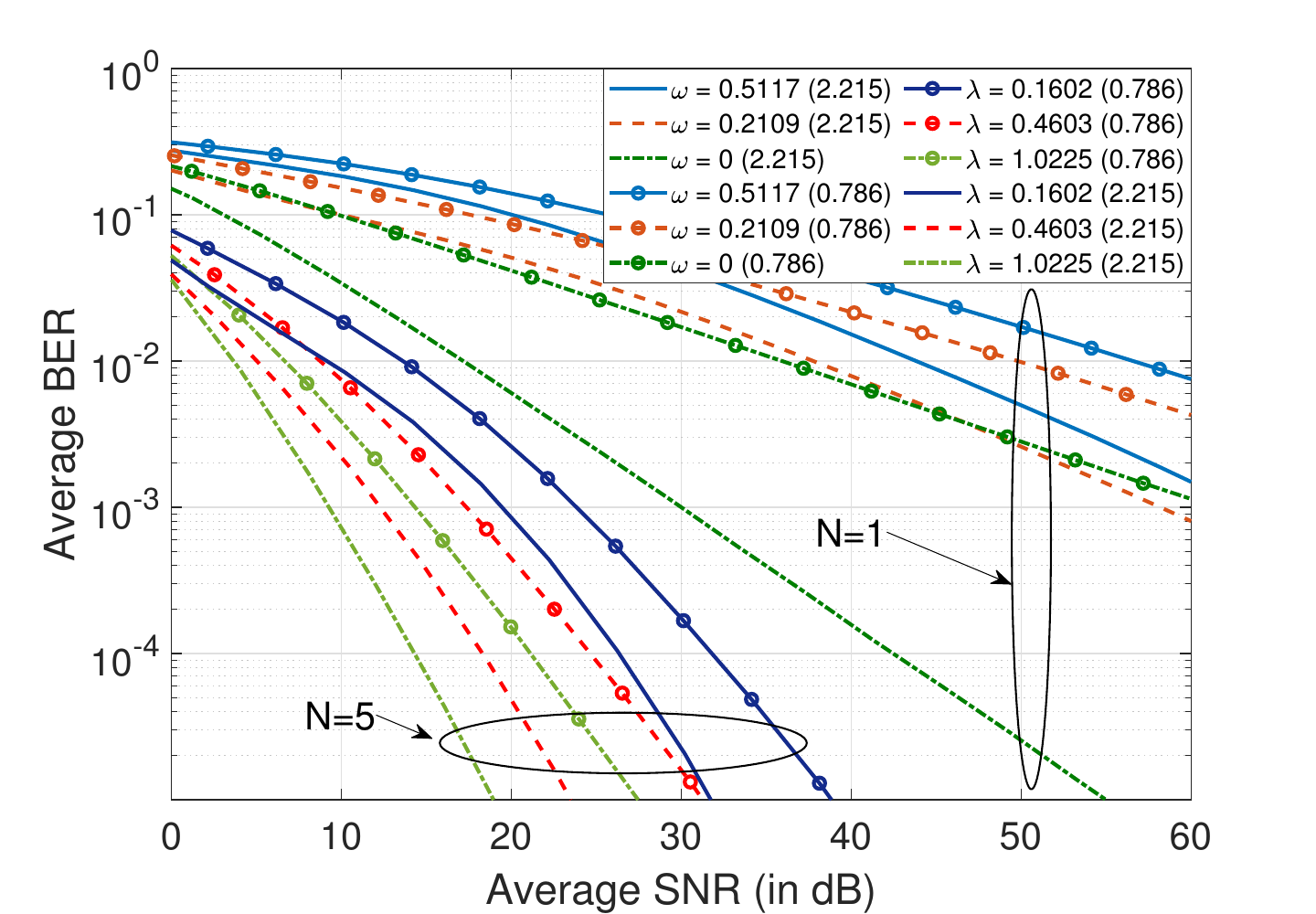}
		\caption{Average BER performance for i.i.d. with $N=1$ and $N=5$ at various values of $\omega$, $\lambda$, and $\rho$. The bracketed values in the legend indicate the values of $\rho$.}
		\label{fig:aber_rho_inid}
	\end{figure}
	
	\begin{figure}
		{\includegraphics[scale=0.395]{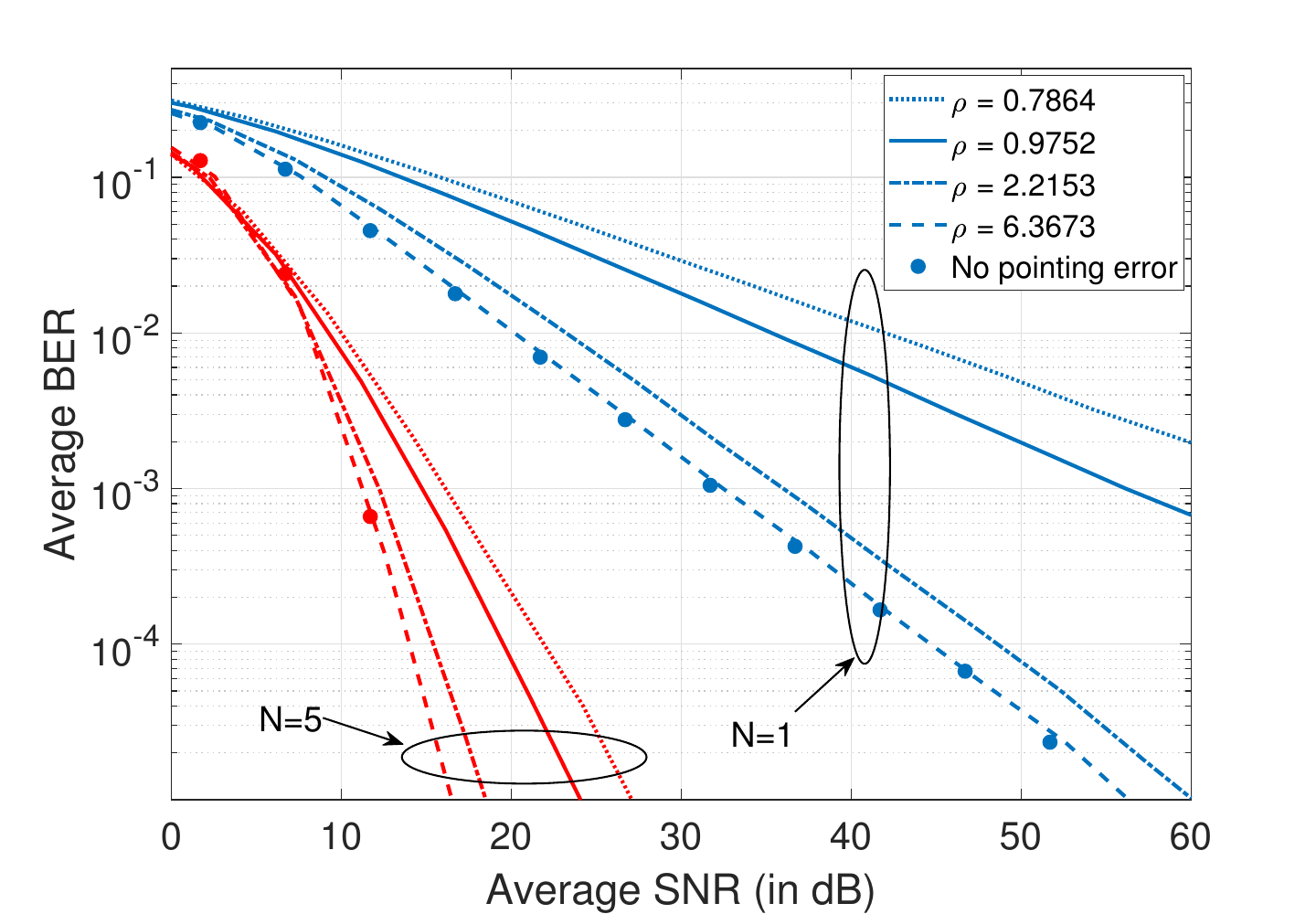}}
		\caption{Effect of pointing errors on the average BER performance for UOWC system.}
		\label{fig:aber_rho_var}
	\end{figure}
	
	\begin{figure*}
		\centering
		\subfigure[Strong turbulence with at $A_0=0.3900$ and $\rho=0.5718$.] {\includegraphics[scale=0.30]{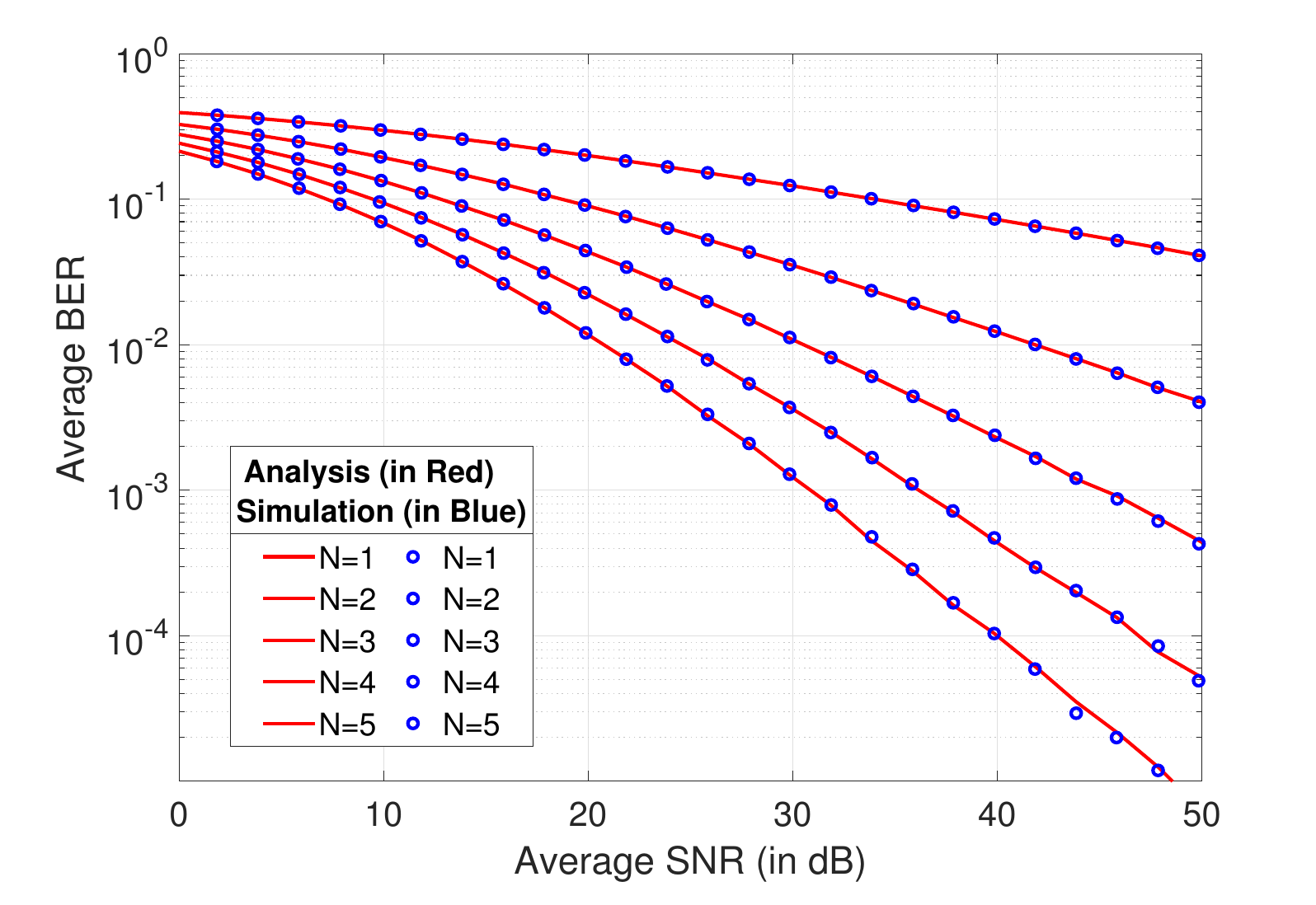} \label{fig:aber_strong}} 
		\subfigure[Weak turbulence with  $A_0=0.3900$ and $\rho=0.5718$.]{\includegraphics[scale=0.30]{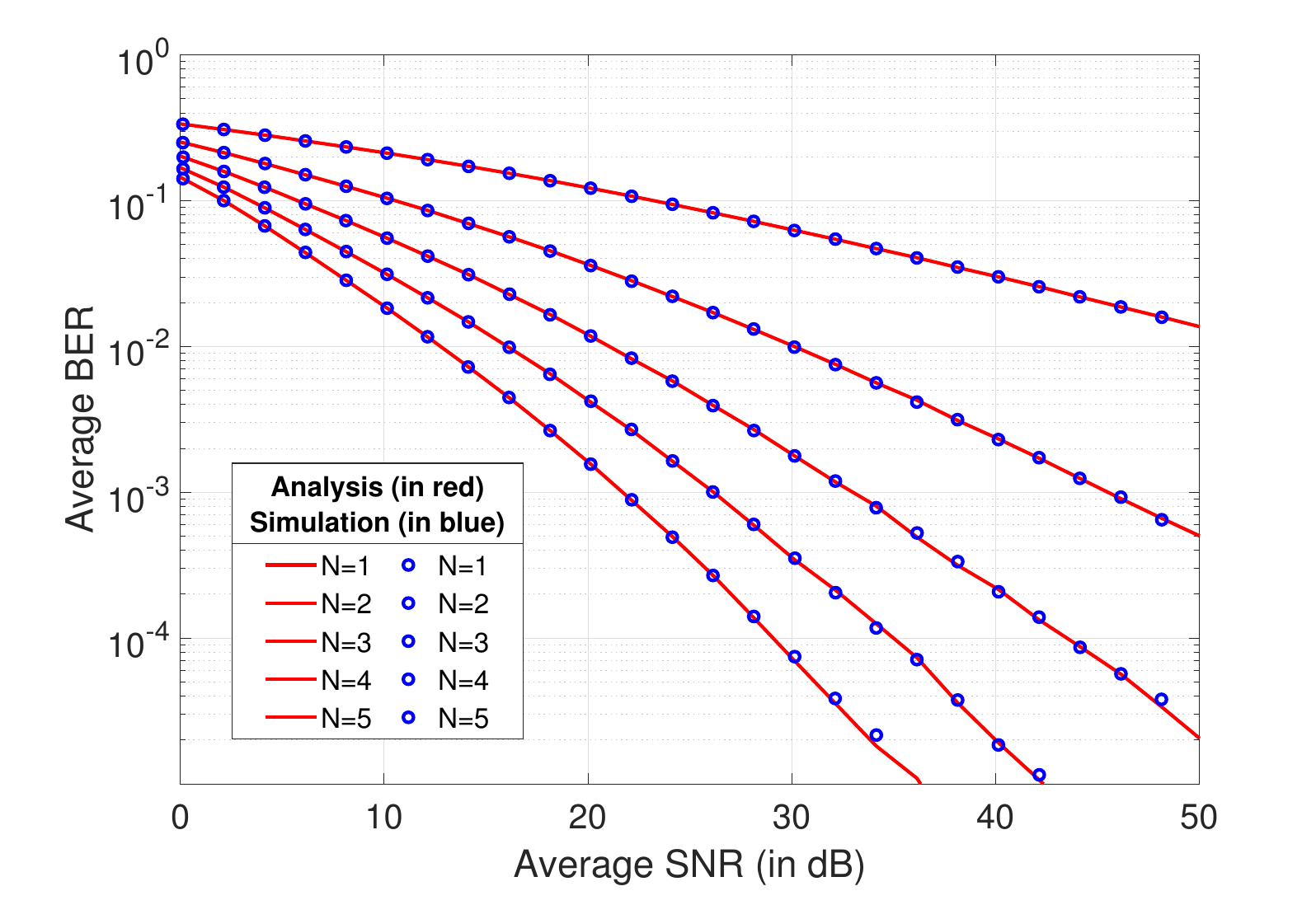}\label{fig:aber_weak}}
		\subfigure[Strong turbulence with negligible pointing errors with $\omega =0$.]{\includegraphics[scale=0.37]{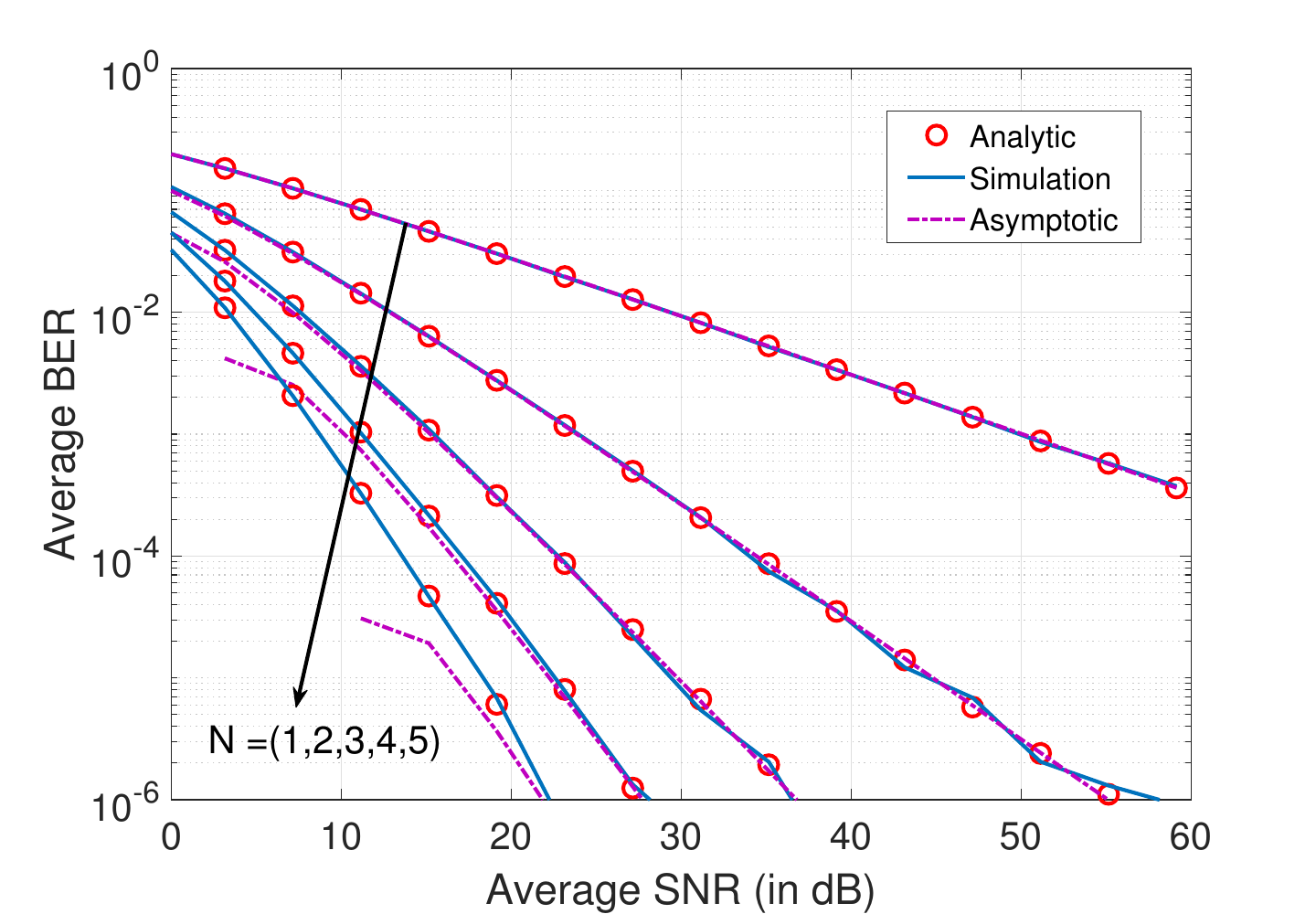}\label{fig:aber_rho_zero}} 
		\caption{Comparing average BER performance for different turbulence conditions and pointing errors.  }
		\label{fig:gradual}
	\end{figure*}

	\section{Simulation Results}

	This section evaluates the average BER and ergodic capacity performance of multi-aperture UOWC system under EGG oceanic turbulence conditions (weak and strong) with and without pointing errors for both i.ni.d. and i.i.d. channel conditions. It should be emphasized that simulation results for i.ni.d. models are rarely available in the literature for UOWC systems, even for more straightforward configurations. We compare the performance of multi-aperture UOWC systems with the single-aperture case ($N = 1$). The analytical expressions derived in this study are validated using Monte-Carlo (MC) simulations averaging over $10^7$ channel realizations. We utilize  MATLAB and Mathematica functions to calculate the Meijer-G and Fox-H functions.

	To simulate weak oceanic turbulence, we consider the set of parameters
	$\{\omega =4.0628 \times 10^{-21}, \lambda_c= 1.0225, a=26.0231, b=0.6993, c=9.5446\}$ and for strong turbulence the set of parameters is $\{\omega =0.5117, \lambda_c= 0.1602, a=0.0075, b=2.9963, c=216.8356\}$ \cite{Zedini2019}. We also consider a particular case of oceanic with zero bubble rate parametrized by $\omega$.   The attenuation of the signal is calculated for an underwater link of $d=50$\mbox{m} as $e^{-\eta d}$, where the extinction coefficient $\eta=0.056$. The pointing error parameters are $A_0=0.1639$ and $\rho=0.9875$ \cite{Farid2007}.

	In Fig.~\ref{fig:aber_inid}, we present the average BER performance of a multiple aperture system under strong turbulence for the i.ni.d. case and compare it with the i.i.d. scenario.  
	To simulate i.ni.d. turbulence for $N=5$, the i.ni.d. scenario is generated by setting all parameters for one link using strong turbulence and applying the same parameters to the other four links, except for one parameter. The varying parameter is selected from four different oceanic turbulence scenarios ranging from weak to strong, as outlined in \cite{Zedini2019}: : $\omega_i\in \{ 4.0628 \times 10^{-21}$, $0.1953$, $0.2109$, $0.3489$\}, $\lambda_i \in \{1.0225$, $0.5273$, $0.4603$, $0.4771$ \}, $a_i\in \{1.40$, $0.7291$, $0.152$, $0.01$\}, $b_i\in \{ 0.6993$, $1.0721$, $1.1501$, $1.4531$\}, $c_i \in \{ 9.5446$, $30.3214$, $41.3258$, $74.3650$\}. The parameters are selected similarly to the $N=5$ case, using the last value from the set of $N=4$  for the  $N=2$ i.ni.d. case.   Comparing the i.i.d. plots for $N=1$, $N=2$, and $N=5$, it can be seen that an increase in the number of apertures improves the average BER performance significantly. The figure shows that an average BER of $10^{-3}$ is not possible with a single aperture system when the average SNR is less than $60$\mbox{dB}. The $N=2$ aperture system achieves an average BER of $10^{-3}$ at an SNR of $43$\mbox{dB}, whereas the $N=5$ aperture system reaches the same BER at an SNR of $26$\mbox{dB}, demonstrating a significant improvement. The i.ni.d. plots show a significant variation in the average BER performance. From the $N=2$ plots, it is evident that varying the values of $a$ and $b$ across the two links has a negligible impact on the average BER performance. However, distinct values of $c$ for the two links increase average BER, while different values of $\lambda$ and $\omega$ have the opposite effect. A similar trend, with more significant variations, is observed in the $N=5$ plots. Thus, the parameter $c$ from a weaker turbulence scenario increases the average BER, while the parameters $\lambda$ and $\omega$ from weaker turbulence decrease the average BER performance.

	In Fig.~\ref{fig:aber_rho_inid}, we consider the i.i.d. scenario to demonstrate the effect of parameters $\omega$ and $\lambda$ on the average BER performance considering low ($\rho=2.215$) and high ($\rho=0.786$) pointing errors. The effect of $\omega$ on the average BER performance is depicted through single aperture $N=1$ plots. It can be seen that the effect of bubble rate parameterized through $\omega$  has a significant effect on the average BER performance:  a lower value of the parameter $\omega$ improves the performance with the lowest $\omega=0$ decreases the average BER more significant. The effect of the parameter $\lambda$ on average BER performance is illustrated through multi-aperture $N=5$ plots. These plots show that an increase in the value of $\lambda$ (indicating weaker turbulence) decreases the average BER performance. In both scenarios, the impact of pointing errors is evident: higher pointing errors reduce the average BER performance, as expected. Comparing  Fig.~\ref{fig:aber_inid} with Fig.~\ref{fig:aber_rho_inid}, it is clear that an average BER of $10^{-3}$ can be achieved within $60$\mbox{dB} even with $N=1$ under certain oceanic turbulence conditions. Both figures demonstrate that the multi-aperture system significantly improves average BER performance. 
	
	In Fig.~\ref{fig:aber_rho_var}, we illustrate the impact of pointing errors on the average BER performance of single and multi-aperture UOWC systems. The results show that lower pointing errors (higher $\rho$ values) lead to a reduced average BER. Notably, the effect of pointing errors is more pronounced in the $N=1$ aperture system compared to the $N=5$ aperture system. At an SNR of approximately $10$ \mbox{dB}, for \( N=1 \) and \( \rho = 0.7864 \), the BER is around \( 10^{-1} \). In contrast, for \( N=5 \) and \( \rho = 6.3673 \), the BER improves significantly to approximately \( 10^{-3} \). At a higher SNR of $40$ \mbox{dB}, the average BER for \( N=1 \) and \( \rho = 0.7864 \) is about \( 10^{-2} \), while for \( N=1 \) and \( \rho = 6.3673 \), it decreases to approximately \( 10^{-4} \). Additionally, in the case of no pointing error, the average BER is further reduced to nearly \( 10^{-5} \) at $40$ \mbox{dB}, emphasizing the substantial impact of pointing errors.

	In Fig.~\ref{fig:gradual},  we depict the gradual effect of increasing the number of apertures ($N=1$ to $5$) on average BER performance under various turbulence conditions and pointing errors. The analysis considers values of $N=1, 2, 3, 4, 5$, strong and weak turbulence, and different pointing errors. A comparison between Fig.~\ref{fig:aber_strong} and Fig.~\ref{fig:aber_weak} reveals that strong turbulence necessitates an additional $10$ \mbox{dB} SNR to achieve the same average BER (e.g., single-aperture systems $N=1$ at $10^{-1}$ average BER and multi-aperture systems $N=5$ at $10^{-4}$ average BER). Increasing the aperture from $N=1$ to $N=2$ enhances performance by reducing the SNR requirement by $10$ \mbox{dB}. In Fig.~\ref{fig:aber_rho_zero}, we examine the scenario with negligible pointing errors and a bubble rate parameter of $\omega=0$. Fig.~\ref{fig:aber_rho_zero} shows a similar trend of improving average BER performance as the number of apertures increases. Thus, The plots in Fig.~\ref{fig:gradual} demonstrate that using more apertures improves BER performance. However, the cumulative gain of the SC-based receiver diminishes as the number of apertures increases, emphasizing the need to optimize the number of apertures to exploit independent channel conditions efficiently. As expected, strong turbulence leads to a higher average BER than weak turbulence. The analytical and asymptotic results are validated through Monte Carlo simulations.
	
\begin{figure*}
		\subfigure[i.ni.d. model]{\includegraphics[scale=0.36]{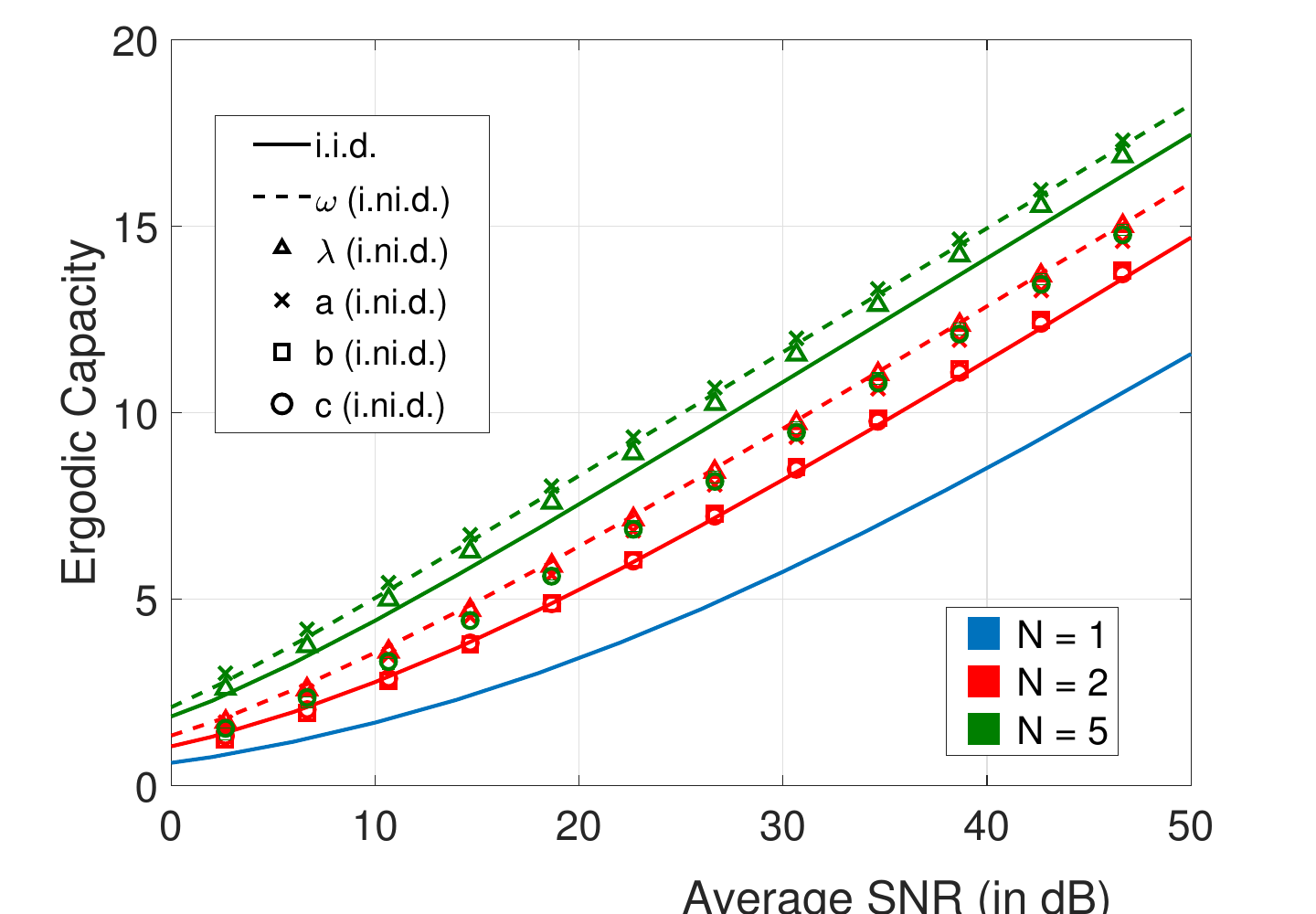}\label{fig:cap_inid}}
		\subfigure[i.i.d. model.]{\includegraphics[scale=0.375]{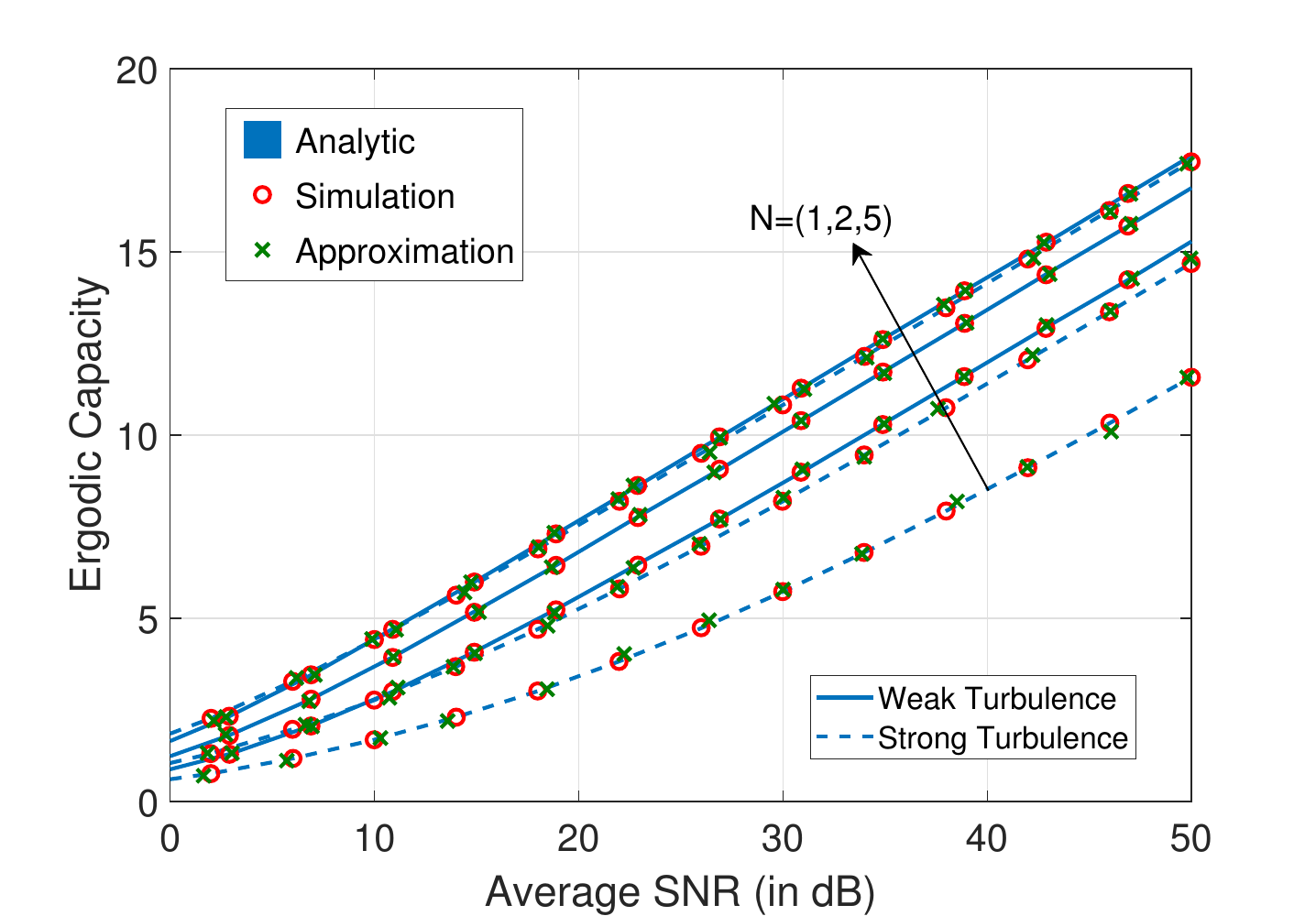}\label{fig:cap_iid}} 
		\caption{Ergodic capacity performance for strong oceanic turbulence conditions with pointing errors  $A_0=0.1639$ and $\rho=0.9875$.}
		\label{fig:capacity}
	\end{figure*}

	Finally, the ergodic capacity performance of the multi-aperture system for both i.i.d. and i.ni.d. models is presented in Fig.~\ref{fig:capacity}. The procedure for generating i.ni.d. turbulence channels follows the same approach described for generating i.ni.d. channels for the average BER performance in  Fig.~\ref{fig:aber_inid}. The bold, markerless plots depict the performance of the i.i.d. channel. The figure shows that increasing the number of apertures enhances the ergodic capacity. At an average SNR of $40$ \mbox{dB}, the ergodic capacity increases from $7$ \mbox{bps/Hz} with $N=1$ to $14$ \mbox{bps/Hz} with $N=5$ apertures. The effect of i.ni.d. channel parameters on ergodic capacity is evident and becomes more significant with an increasing number of apertures. Specifically, weaker turbulence characterized by parameters $\lambda$ and $\omega$ improves performance, while parameter $c$ from weaker turbulence decreases ergodic capacity, and the effects of $a$ and $b$ are minimal. In Fig.~\ref{fig:cap_iid}, weak and strong turbulence scenarios are considered to show the ergodic capacity with increasing apertures. An interesting observation is that increasing the number of apertures reduces the gap between strong and weak turbulence. With $N=1$, the gap is $3$ \mbox{bps/Hz}, but it reduces almost entirely by $N=5$. We validate the derived analytical results through numerical evaluations of exact and approximate analytical expressions using Monte Carlo simulations.

	\section{Conclusion}
	This paper presented a general framework utilizing the continued product and positive integer exponents of Meijer-G function sums to analyze UOWC system performance statistically for both i.ni.d. and i.i.d. channels. We also presented a computationally efficient approach to approximate positive integer exponents of the Meijer-G function using a single-variate Fox-H function. We derived exact and approximate expressions for the average BER and ergodic capacity under the combined effect of the EGG turbulence and pointing errors. We conducted extensive simulations for average BER and ergodic capacity, considering both i.ni.d. and i.i.d. models to assess the impact of turbulence and pointing error parameters. Simulation results highlighted that the exact expressions provide a more reliable estimate for the efficient deployment of UOWC systems. Simulations demonstrated that the low-complexity SC technique significantly improves performance under challenging ocean turbulence and pointing errors by utilizing spatial diversity. Further,  analysis of the i.ni.d. model revealed that the EGG turbulence parameter $c$ from weaker turbulence (low values of $c$) increases the average BER. The parameters $\lambda$ and $\omega$  from weaker turbulence (high values of $\lambda$ and low values of $\omega$) have the opposite effect. Increasing the number of apertures mitigates the impact of strong turbulence, resulting in similar ergodic capacity across strong and weaker turbulence conditions. Experimental validation of the proposed analysis offers a valuable avenue for future research.
	
	\section{Acknowledgment}
	The authors would like to thank Mr. Syed Maaz Husain (UG student, BITS Pilani) and Mr. Piyush Sial (UG student, BITS Pilani) for their help in the conference version of this paper.
	\section*{Appendix A: Proof of Lemma 1}
	Using \eqref{cdf_uw_pe} and \eqref{cdf_sc_inid}  in  equation \eqref{BER1}, we get
	\begin{flalign}\label{cdf_inid_pe4}
		&\bar{P}_{e} = \frac{q^p\rho^{2N}}{2\Gamma(p)} \int_{0}^{\infty} e^{-q\gamma} \gamma^{p-1} \nonumber \\& \times
		\prod_{i = 1}^{N} \bigg[ \omega_i G_{2,3}^{2,1}\left(\begin{array}{c}1,\rho_i^2+1\\1,\rho_i^2,0\end{array}\left|\frac{1}{\lambda_i A_i}\left(\sqrt{\frac{\gamma}{\bar{\gamma}_i}}\right)\right.\right)   \nonumber \\& +
		\frac{1-\omega_i}{c_i\Gamma(a_i)}G_{2,3}^{2,1}\left(\begin{array}{c}1,\frac{\rho_i^2}{c_i}+1\\a_i,\frac{\rho_i^2}{c_i},0\end{array}\left|\frac{1}{b_i^{c_i} A_i^{c_i}}\left(\sqrt{\frac{\gamma}{\bar{\gamma}_i}}\right)^{c_i}\right.\right) \bigg]
		d\gamma
	\end{flalign}
	Applying  \eqref{eq:IN12} from  Proposition \ref{contproduct_sum}, we can express  \eqref{cdf_inid_pe4} as
	\begin{flalign}\label{cdf_inid_pe5}
		&\bar{P}_{e} = \frac{q^p\rho^{2N}}{2\Gamma(p)} \int_{0}^{\infty} e^{-q\gamma} \gamma^{p-1} \nonumber \\& \times
		\sum_{S \subseteq \{1,2,.., N\}} \bigg [ \prod_{i \epsilon S} \omega_i G_{2,3}^{2,1}\left(\begin{array}{c}1,\rho_i^2+1\\1,\rho_i^2,0\end{array}\left|\frac{1}{\lambda_i A_i}\left(\sqrt{\frac{\gamma}{\bar{\gamma}_i}}\right)\right.\right) \nonumber \\ & \times \prod_{j \epsilon S^C} \frac{1-\omega_j}{c_j\Gamma(a_j)}G_{2,3}^{2,1}\left(\begin{array}{c}1,\frac{\rho_j^2}{c_j}+1\\a_j,\frac{\rho_j^2}{c_j},0\end{array}\left|\frac{1}{b_j^{c_j} A_j^{c_j}}\left(\sqrt{\frac{\gamma}{\bar{\gamma}_j}}\right)^{c_j}\right.\right) \bigg ]
		d\gamma
	\end{flalign}
	Using the definition of Meijer-G function in \eqref{eq:mg_def} in \eqref{cdf_inid_pe5}, and  applying Fubini's theorem:
	\begin{flalign}\label{cdf_inid_pe7}
		&\bar{P}_{e} = \frac{q^p\rho^{2N}}{2\Gamma(p)}
		\sum_{S \subseteq \{1,..,N\}}  \oint \limits_{L_1} \cdots \oint \limits_{L_N}  
		\prod_{i \epsilon S} \left( \omega_i   F(s_i) (\psi_i)^{s_i} \right) \nonumber \\ & \times \prod_{j \epsilon S^c} \left( \frac{1-\omega_j}{c_j\Gamma(a_j)}  F(s_j) (\psi_j )^{s_j} \right)   \nonumber \\& \times \left( \int_{0}^{\infty} e^{-q\gamma} \gamma^{\left(\frac{1}{2}(\sum_{S}s_i + \sum_{S^c}c_js_j\right) +p-1} d\gamma \right) ds_1 \cdots ds_N
	\end{flalign}
	Denoting 	 $\psi_i = \frac{1}{\lambda_i A_i}\left(\sqrt{\frac{1}{\bar{\gamma}_i}}\right)$,	 $\psi_j = \frac{1}{b_j^{c_j} A_j^{c_j}}\left(\sqrt{\frac{\gamma}{\bar{\gamma}_j}}\right)^{c_j} $,  solving $\int_{0}^{\infty} e^{-q\gamma} \gamma^{\left(\frac{1}{2}(\sum_{S}s_i + \sum_{S^c}c_js_j\right) +p-1} d\gamma= \Gamma \left(p+\frac{1}{2}\left(\sum_{S}s_i + \sum_{S^c}c_js_j\right)\right)$, and applying the definition of multivariate Fox-H function \cite{mathai2009h}, we get \eqref{cdf_inid_pe10}.

	\section*{Appendix B: Proof of Lemma 4}
	
	Using \eqref{eq:cap_inid_f1_zaf}  in \eqref{eq:cap_eq}, we express the ergodic capacity as  $C=C_1+C_2$, where
	\begin{flalign}\label{eq:cap_inid_f3_zaf}
		&C_1=\int_{0}^\infty\log_2(1+\gamma) \sum_{k =1}^{N} \frac{\omega_k \rho_k^{2n} }{2 \gamma} \nonumber \\& \times \prod_{n \neq k} F_{X_n}(x)   G_{1,2}^{2,0}\left(\begin{array}{c}\rho_k^2+1\\1,\rho_k^2\end{array}\left|\frac{1}{\lambda_k A_k} \left(\sqrt{\frac{\gamma}{\bar{\gamma}}}\right)\right.\right) d\gamma\nonumber \\ & 
	\end{flalign}
	\begin{flalign}\label{eq:cap_inid_f3_zaf2}
		&C_2=\int_{0}^\infty\log_2(1+\gamma) \frac{(1-\omega_k)\rho_k^{2n}}{2 \Gamma(a_k) \gamma} \prod_{n \neq k} \left(F_{X_n}(x)\right) \nonumber \\& \times  G_{1,2}^{2,0}\left(\begin{array}{c}\frac{\rho_k^2}{c_k}+1\\a_k,\frac{\rho_k^2}{c_k}\end{array}\left|\frac{1}{b_k^{c_k} A_k^{c_k}}\left(\sqrt{\frac{\gamma}{\bar{\gamma}_k}}\right)^{c_k}\right.\right) d\gamma 
	\end{flalign}

	Applying product of continued sum (i.e., Proposition \ref{contproduct_sum}) in \eqref{cdf_uw_pe}   to express:
	\begin{flalign}\label{eq:cap_inid_f3}
		&\prod_{n=1}^{N} F_{X_n}(x) = \prod_{i = 1}^N \rho_i^{2}\sum_{S} \bigg [ \prod_{i \epsilon S} \omega_i  \nonumber \\ & \times G_{2,3}^{2,1}\left(\begin{array}{c}1,\rho_i^2+1\\1,\rho_i^2,0\end{array}\left|\frac{1}{\lambda_i A_i}\left(\sqrt{\frac{\gamma}{\bar{\gamma}_i}}\right)\right.\right) \nonumber \\ & \times \prod_{j \epsilon S^C} \frac{1-\omega_j}{c_j\Gamma(a_j)}G_{2,3}^{2,1}\left(\begin{array}{c}1,\frac{\rho_j^2}{c_j}+1\\a_j,\frac{\rho_j^2}{c_j},0\end{array}\left|\frac{1}{b_j^{c_j} A_j^{c_j}}\left(\sqrt{\frac{\gamma}{\bar{\gamma}_j}}\right)^{c_j}\right.\right) \bigg ]
	\end{flalign}	
	We require the Meijer-G representation of the following function:	
	\begin{flalign}\label{eq:log_meijer_1}
		\frac{\ln(1+\gamma)}{\gamma} = G_{2,2}^{1,2}\left(\begin{array}{c}0,0\\0, -1 \end{array}\bigg|  \gamma \right)
	\end{flalign}
	
	Using \eqref{eq:cap_inid_f3} in \eqref{eq:cap_inid_f3_zaf} with \eqref{eq:log_meijer_1}, we get 
	\begin{flalign}
		& C_1 = \sum_{k =1}^{N} \frac{\omega_k \prod_{n = 1}^N (\rho_n^{2}) }{2 \ln(2)} \int_{0}^{\infty} G_{2,2}^{1,2}\left(\begin{array}{c}0,0\\0, -1 \end{array}\bigg|  \gamma \right) \nonumber \\ & \times  \sum_{S \subseteq \{1,2,.., N \}} \bigg [ \prod_{i \epsilon S} \omega_i G_{2,3}^{2,1}\left(\begin{array}{c}1,\rho_i^2+1\\1,\rho_i^2,0\end{array}\left|\frac{1}{\lambda_i A_i}\left(\sqrt{\frac{\gamma}{\bar{\gamma}_i}}\right)\right.\right) \nonumber \\ & \times \prod_{j \epsilon S^C} \frac{1-\omega_j}{c_j\Gamma(a_j)}G_{2,3}^{2,1}\left(\begin{array}{c}1,\frac{\rho_j^2}{c_j}+1\\a_j,\frac{\rho_j^2}{c_j},0\end{array}\left|\frac{1}{b_j^{c_j} A_j^{c_j}}\left(\sqrt{\frac{\gamma}{\bar{\gamma}_j}}\right)^{c_j}\right. \right) \bigg ] \nonumber \\ & \times G_{1,2}^{2,0}\left(\begin{array}{c}\rho_k^2+1\\1,\rho_k^2\end{array}\left|\frac{1}{\lambda_k A_k} \left(\sqrt{\frac{\gamma}{\bar{\gamma}_k}}\right)\right.\right)  d\gamma
	\end{flalign}
	Using the integral representation of the Meijer-G function with the solution of the inner integral in $\gamma$, we get
	\begin{flalign}\label{eq:cap_inid_c1_int}
		& C_1 = \sum_{k =1}^{N} \frac{\omega_k\prod_{n = 1}^N (\rho_n^{2}) }{ln(2)} \lambda_k^2 A_k^2 \bar{\gamma_{k}} \oint\limits_{L_0} \oint\limits_{L_1}..\oint\limits_{L_N} F_1(s_0) \left(\lambda_k^2 A_k^2 \bar{\gamma_{k}} \right)^{s_0}  \nonumber \\ & \times \sum_{S} \bigg [ \prod_{i \epsilon S} \omega_i F_{2i}(s_{i}) \left( \frac{\lambda_k A_k}{\lambda_i A_i}\sqrt{\frac{\bar{\gamma}_k}{\bar{\gamma}_i}}\right)^{s_i}  \nonumber \\ & \times \prod_{j \in S^c} \frac{1-\omega_j}{c_j\Gamma(a_j)}F_{2j}(S_{j})  \left(\frac{\lambda_k A_k}{b_j A_j} \sqrt{\frac{\bar{\gamma}_k}{\bar{\gamma}_j}}\right)^{s_jc_j} \bigg ]  \nonumber \\ & \times \frac{\Gamma(3+x_1) \Gamma(\rho_k^2+2+x_1)}{\Gamma(3+\rho_k^2+x_1)} ds_0ds_1..ds_N
	\end{flalign}
	where, $x_1 = 2s_0+\sum_{S} s_i + c\sum_{S^c} s_j$
	Applying the definition of multivariate Fox-H function in \eqref{eq:cap_inid_c1_int}, we get a closed-form expression for $C_1$, which is the first of  \eqref{eq:cap_inid_exact}. Applying the similar method for $C_2$, we get the second term of  \eqref{eq:cap_inid_exact}, which completes the proof.
	
	\section*{Appendix C: Proof of Lemma 5}
	Substituting the Meijer-G function representation of  logarithmic function \eqref{eq:log_meijer_1} in \eqref{eq:cap_iid_2}, we get
	\begin{flalign}\label{eq:cap_iid_3}
		& C =    \frac{Nc}{2\ln(2)} \left(\frac{\rho^2}{c \Gamma(a)}\right)^N  \int_{0}^{\infty} G_{2,2}^{1,2}\left(\begin{array}{c}0,0\\0, -1 \end{array}\bigg|  \gamma \right)   \nonumber \\ & \times \bigg [G_{2,3}^{2,1}\left(\begin{array}{c}1,\frac{\rho^2}{c}+1\\a,\frac{\rho^2}{c},0\end{array}\left|\frac{1}{b^{c} A^{c}}\left(\sqrt{\frac{\gamma}{\bar{\gamma}}}\right)^{c}\right.\right) \bigg ]^{N-1} \nonumber \\ & \times G_{1,2}^{2,0}\left(\begin{array}{c}\frac{\rho^2}{c}+1\\a,\frac{\rho^2}{c}\end{array}\left|\frac{1}{b^{c} A^{c}}\left(\sqrt{\frac{\gamma}{\bar{\gamma}}}\right)^{c}\right.\right) d\gamma
	\end{flalign}
	Applying Proposition \ref{prop1}, we express \eqref{eq:cap_iid_3}
	\begin{flalign}\label{eq:cap_iid_5}
		& C =    \frac{Nc}{2\ln(2)} \left(\frac{\omega\rho^2}{c \Gamma(a)}\right)^N  \oint \limits_{L_0} \oint \limits_{L_1} \cdots \oint \limits_{L_{N-1}}  F(s_0)  F(s_1)  \cdots \nonumber \\ & \times \left( \frac{1}{bA}\sqrt{\frac{\gamma}{\bar{\gamma}}} \right)^{cs_1} \cdots F(s_{N-1}) \left( \frac{1}{bA}\sqrt{\frac{\gamma}{\bar{\gamma}}} \right)^{cs_{N-1}}   \nonumber \\ & \times \int\limits_{0}^{\infty} \gamma^{s_0}  \gamma^{ \frac{c(\sum_{i=1}^{N-1}s_i)}{2}}  G_{1,2}^{2,0}\left(\begin{array}{c}\frac{\rho^2}{c}+1\\a,\frac{\rho^2}{c}\end{array}\left|\frac{1}{b^{c} A^{c}}\left(\sqrt{\frac{\gamma}{\bar{\gamma}}}\right)^{c}\right.\right)  \nonumber \\ & \times d\gamma ds_0 ds_1\cdots ds_{N-1}
	\end{flalign}
	Solving the inner integral in terms of Gamma function \cite{Wolfram_meijer} and applying the multivariate Fox-H definition, we get \eqref{eq:cap_iid_2}.
	\section*{Appendix D: Proof of Lemma 6}
	Using \eqref{eq:log_meijer_1} in \eqref{eq:cap_eq}, we can get
	\begin{flalign}\label{cap4b}
		& C =    \frac{Nc}{2\ln(2)} \left(\frac{\omega\rho^2}{c \Gamma(a)}\right)^N  \int_{0}^{\infty} \oint \limits_{L_1}  \oint \limits_{L_2} F(s_1)  \gamma^{s_1} F(s_2)   \nonumber \\ & \times \bigg [G_{2,3}^{2,1}\left(\begin{array}{c}1,\frac{\rho^2}{c}+1\\a,\frac{\rho^2}{c},0\end{array}\left|\frac{1}{b^{c} A^{c}}\left(\sqrt{\frac{1}{\bar{\gamma}}}\right)^{c}\right.\right) \bigg ]^{N-2} 	\left( \frac{1}{bA}\sqrt{\frac{\gamma}{\bar{\gamma}}}\right)^{cs_2 (N-1)}\nonumber \\ & \times
		G_{1,2}^{2,0}\left(\begin{array}{c}\frac{\rho^2}{c}+1\\a,\frac{\rho^2}{c}\end{array}\left|\frac{1}{b^{c} A^{c}}\left(\sqrt{\frac{\gamma}{\bar{\gamma}}}\right)^{c}\right.\right) d\gamma ds_1 ds_2
	\end{flalign}
	Applying  Proposition \ref{prop3} in \eqref{cap4b}, we get
	\begin{flalign}\label{eq:iid_cap_int_foxh}
		& C =    \frac{N} {\ln(2)} \left(\frac{\omega\rho^2}{c \Gamma(a)}\right)^N  \oint \limits_{L_1}  \oint \limits_{L_2} F(s_1)   F(s_2)  \bigg(\frac{1}{bA\sqrt{\gamma_{0}}}\bigg)^{-2s_1 -2 }  \nonumber \\ & \times \bigg [G_{2,3}^{2,1}\left(\begin{array}{c}1,\frac{\rho^2}{c}+1\\a,\frac{\rho^2}{c},0\end{array}\left|\frac{1}{b^{c} A^{c}}\left(\sqrt{\frac{1}{\bar{\gamma}}}\right)^{c}\right.\right) \bigg ]^{N-2} \nonumber \\ & \times \frac{\Gamma (a+ \frac{2s_1}{c}+ \frac{2}{c} +  (N-1)s_2 )}{\Gamma(1+ \frac{\rho^2}{c} + \frac{2s_1}{c}+ \frac{2}{c} +  (N-1)s_2 )}  \nonumber \\ & \times  \Gamma \left(\frac{\rho^2}{c} + \frac{2s_1}{c}+ \frac{2}{c} +  (N-1)s_2\right) ds_1 ds_2
	\end{flalign}
	A straightforward application of the definition of the bivariate function, we get \eqref{cap_final_iid}, which completes the proof.
	
	\bibliographystyle{IEEEtran}
	\bibliography{bib_file}

\end{document}